\documentclass[prper,aps,twocolumn,groupedaddress,floats,final,superscriptaddress]{revtex4-1}
\usepackage{graphicx}
\usepackage{dcolumn}
\usepackage{bm}
\usepackage{color}
\usepackage{amsmath}
\usepackage{amssymb}
\usepackage{amsfonts}
\usepackage{enumitem}
\RequirePackage[
  hyperindex,colorlinks,bookmarksnumbered,
  plainpages=true,pdfstartview=FitH]{hyperref}
\hypersetup{linkcolor=blue,urlcolor=blue,citecolor=blue}
\usepackage{appendix}
\usepackage{hyperref}

\AtBeginDocument{%
    \newwrite\bibnotes
    \def\bibnotesext{Notes.bib}
    \immediate\openout\bibnotes=\jobname\bibnotesext
    \immediate\write\bibnotes{@CONTROL{REVTEX41Control}}
    \immediate\write\bibnotes{@CONTROL{%
    apsrev41Control,author="08",editor="1",pages="1",title="0",year="1"}}
     \if@filesw
     \immediate\write\@auxout{\string\citation{apsrev41Control}}%
    \fi
}%

\newcommand{\di}{\mathrm{d}}
\newcommand{\gi}{\mathrm{g}}
\newcommand{\ci}{\mathrm{c}}

\newcommand{\beq}{\begin{equation}}
\newcommand{\eeq}{\end{equation}}
\newcommand{\ba}{\begin{array}}
\newcommand{\ea}{\end{array}}
\newcommand{\bea}{\begin{eqnarray}}
\newcommand{\eea}{\end{eqnarray}}

\begin{document}

\title{Supervised machine learning of ultracold atoms with speckle disorder}

\author{S. Pilati}
\affiliation{School of Science and Technology, Physics Division, Universit{\`a}  di Camerino, 62032 Camerino (MC), Italy}
\author{P. Pieri}
\affiliation{School of Science and Technology, Physics Division, Universit{\`a}  di Camerino, 62032 Camerino (MC), Italy}
\affiliation{INFN, Sezione di Perugia, 06123 Perugia (PG), Italy}

\begin{abstract}
We analyze how accurately supervised machine learning techniques can predict the lowest energy levels of one-dimensional noninteracting ultracold atoms
subject to the correlated disorder due to an optical speckle field.
Deep neural networks with different numbers of hidden layers and neurons per layer are trained on large sets of instances of the speckle field, whose energy levels have been preventively determined via a high-order finite difference technique. The Fourier components of the speckle field are used as feature vector to represent the speckle-field instances. A comprehensive analysis of the details that determine the possible success of supervised machine learning tasks, namely the depth and the width of the neural network, the size of the training set, and the magnitude of the regularization parameter, is presented. It is found that ground state energies of previously unseen instances can be predicted with essentially arbitrary accuracy. First and second excited state energies can be predicted too, albeit with slightly lower accuracy and using more layers of hidden neurons.
\end{abstract}

\maketitle
\section{Introduction}
Machine learning techniques are at the heart of various technologies used in every day life, like e-mail spam filtering, voice recognition software, and web-text analysis tools.
They have already acquired relevance also in physics and chemistry research. In these fields, they have been employed for diverse tasks such as, e.g., finding energy-density functionals~\cite{snyder2012finding,li2016understanding,brockherde2017bypassing,snyder2013orbital}, identifying phases and phase transitions in many-body systems~\cite{wang2016discovering,carrasquilla2017machine,van2017learning,ch2017machine,wetzel2017unsupervised,deng2017machine,ohtsuki2017deep}, predicting properties such as the atomization energy of molecules and crystals from large databases of known compounds~\cite{hansen2013assessment,hansen2015machine,schutt2014represent}, or predicting ligand-protein poses and affinities for drug-design research~\cite{ragoza2017protein,wojcikowski2017performance,khamis2015machine,pereira2016boosting}.
In particular, supervised machine learning has been put forward as a fast, and possibly accurate, technique to predict the energies of quantum systems exploiting the information contained in large datasets of data obtained using computationally expensive numerical tools~\cite{mills2017deep}. This has already proven to be a very promising approach to  determine the potential energy surfaces for molecular dynamics simulations of materials, of chemical compounds, and of biological systems, providing a boost in speed and in accuracy~\cite{blank1995neural,lorenz2004representing,behler2007generalized,handley2010potential,bartok2010gaussian,behler2011neural,behler2011atom}.
However, it is not yet precisely known how accurately the statistical models commonly employed in supervised machine learning can describe quantum systems. In general, the accuracy achievable by these statistical models, chiefly artificial neural networks, depends on various important details, including the depth and the connectivity structure of the neural network, the size of the training set, and the type of regularization employed during the training process~\cite{lecun2015deep}. Also the choice of the features adopted to represent the quantum system of interest plays a crucial role; in fact, a lot of research work has been devoted to the development of efficient representations (see, e.g., Refs.~\cite{behler2016perspective,khamis2015machine}).
It is natural to expect that addressing models that describe highly tunable and easily accessible experimental setups could shed some light on this important issue.\\

These considerations lead us to focus on ultracold atom experiments. In fact, these systems have emerged in recent years as an ideal platform to investigate quantum many-body phenomena~\cite{giorgini2008theory,bloch2008many}. They allowed experimentalists to implement archetypal Hubbard-type models of condensed matter physics~\cite{jaksch2005cold} and even the realization of programmable simulators of quantum spin Hamiltonians~\cite{bernien2017probing}. One of the quantum phenomena that received most consideration in this research area is the Anderson localization transition in the presence of disorder~\cite{roati2008anderson,billy2008direct,aspect2009anderson,kondov2011three,jendrzejewski2012three}. This phenomenon consists in the spatial localization of the single particle states, determining the absence of transport in macroscopic samples~\cite{anderson1958absence}. Unlike conventional condensed matter systems, which inherently include a certain amount of impurities, in cold-atom setups disorder is introduced on purpose. The most frequently used technique consists in creating optical speckle fields by shining lasers through rough semitransparent membranes, and then focusing them onto the atomic cloud. These speckle fields are characterized by a particular structure of the spatial autocorrelation of the local optical field intensities~\cite{goodman1975statistical,goodman2007speckle}. 
These correlations have to be accounted for in the modeling of cold-atom experiments with speckle fields~\cite{falco2010density,modugno2010anderson}. Indeed, they determine the position of the mobility edge~\cite{delande2014mobility,fratini,fratini2015anderson}, namely the energy threshold that in three dimensional systems separates the localized states from the extended ergodic states. In low dimensional configurations, any amount of disorder is sufficient to induce Anderson localization. However, the speckle-field correlations determine the transport properties and even the emergence of so-called effective mobility edge, i.e. energy thresholds where the localization length changes abruptly~\cite{PhysRevLett.82.4062,PhysRevLett.98.210401,PhysRevA.80.023605}.\\

In this article we perform a supervised machine learning study of the lowest three energy levels of a one-dimensional quantum particle moving in a disordered external field. This model is designed to describe an alkali atom exposed to a one-dimensional optical speckle filed, taking into account the detailed structure of the spatial correlations of the local intensities of the speckle field. This is in fact the setup implemented in the first cold-atom experiments on Anderson localization~\cite{roati2008anderson,billy2008direct}.
First, we determine the energy levels of a large set of speckle-field instances via a high-order finite difference formula. Next, we train a deep artificial neural network to reproduce the energy levels of this training set, and we then test how accurately the neural network predicts the energy levels of previously unseen speckle-field instances.
We analyze in detail how the prediction accuracy varies with the most relevant parameters that influence the success of supervised machine learning tasks, in particular the size of the training set, the number of hidden layers in the neural network, the number of neurons per hidden layer, and the magnitude of the regularization parameter.
The main result is that, given a sufficiently large training set, a neural network with several layers can predict ground state energies with essentially arbitrary accuracy.
Higher energy levels can be predicted too, but the accuracy of these predictions is slightly lower and requires the training of deeper neural networks.
%

%
\begin{figure}
\label{fig1}
\begin{center}
\includegraphics[width=1.0\columnwidth]{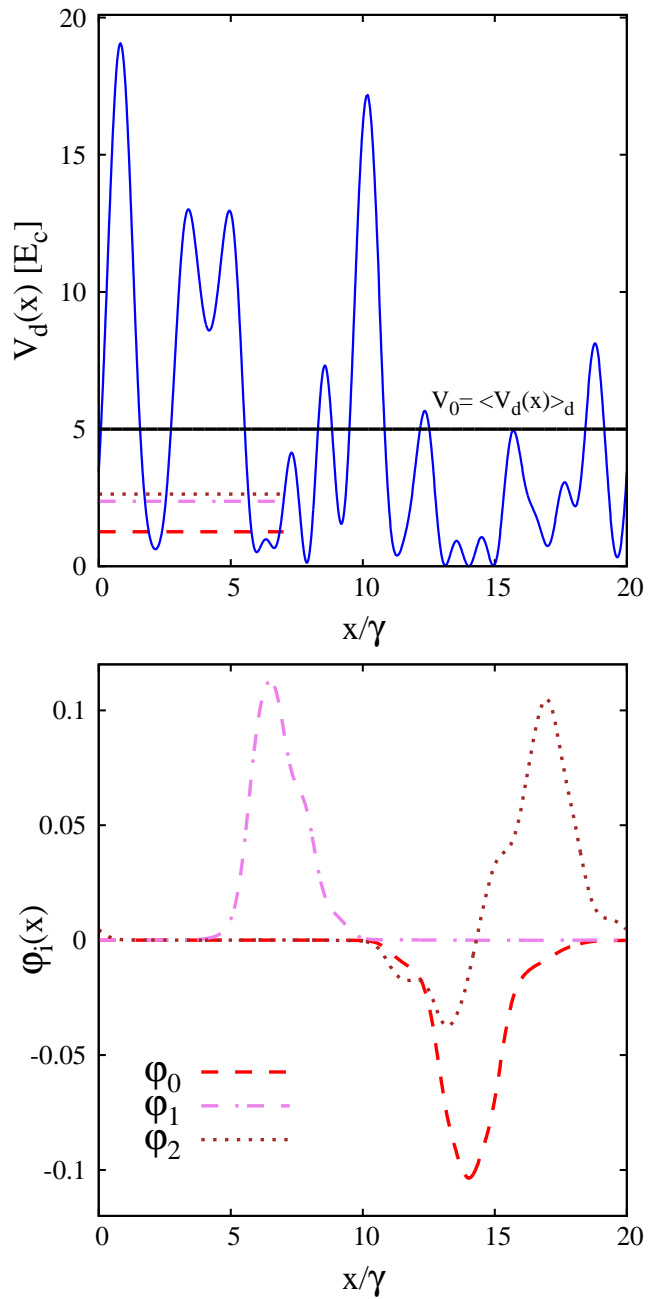}
\caption{(color online). 
Upper panel: local intensity $V_{\di}(x)$ of a typical instance of optical speckle field, as a function of the spatial coordinate $x/\gamma$. The energy unit is the correlation energy $E_{\ci}= \hbar^2/\left(2m\gamma^2\right)$, defined by the spatial correlation length of the speckle field $\gamma$.
The continuous horizontal (black) line indicates the average over many speckle-field instances of the optical speckle field intensity $V_0=\langle V_d(x)\rangle_d$.
Lower panel: profile of the three lowest-energy wave functions of the speckle field instance displayed in the upper panel. The corresponding energy levels are indicated by the three horizontal segments in the upper panel.
}

\end{center}
\end{figure}
\begin{figure}
\begin{center}
\includegraphics[width=1.0\columnwidth]{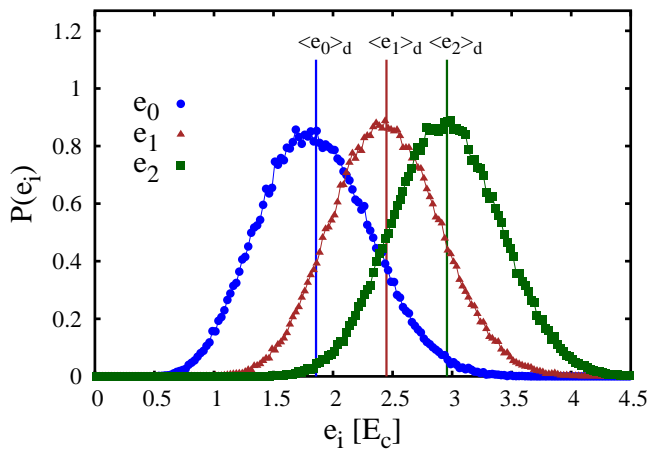}
\caption{(color online). 
Probability distribution $P(e_i)$ of the first three energy levels $e_0$, $e_1$, and $e_2$. The vertical segments indicate the corresponding averages over many speckle-field instances  $\left<e_0\right>_{\di}$, $\left<e_1\right>_{\di}$, and $\left<e_2\right>_{\di}$. The energy unit is the correlation energy $E_{\ci}$.
}
\label{fig2}
\end{center}
\end{figure}

\begin{figure}
\begin{center}
\includegraphics[width=1.0\columnwidth]{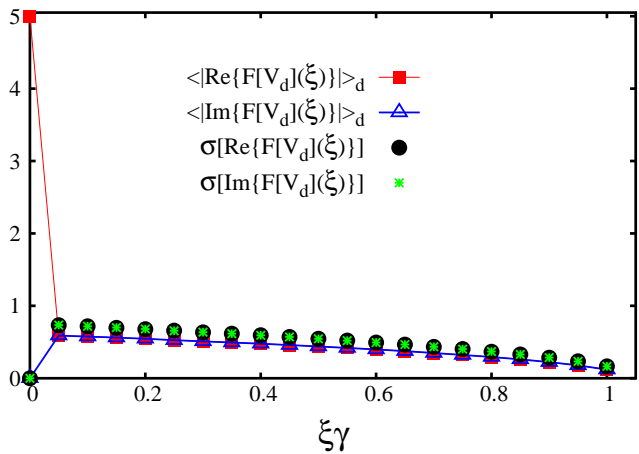}
\caption{(color online). 
Average over many speckle-field instances of the absolute value of the Fourier components of the speckle field. The (red) squares correspond to the real part. 
The (blue) open triangles correspond to the imaginary part. The (black) circles and the (green) stars indicate the standard deviations of the real part and of the imaginary part, respectively.
}
\label{fig3}
\end{center}
\end{figure}

\section{Ultracold atoms in a one-dimensional optical speckle field}
\label{secspeckle}
The model we consider is defined by a Hamiltonian operator that in coordinate representation reads:
\begin{equation}
\hat{H}=-\frac{\hbar^2}{2m} \frac{\mathrm{d}^2}{\mathrm{d}x^2} + V_{\di}(x),
\label{H}
\end{equation}
where $\hbar$ is the reduced Planck's constant and $m$ the particle mass. $V_{\di}(x)$ is a disordered external field, designed to represent the potential energy of an atom subject to an optical speckle field.
Experimentally, these optical fields are generated when coherent light passes through, or is reflected by, rough surfaces. In the far field regime, a specific light intensity pattern develops, commonly referred to as optical speckle field. In cold-atom experiments, this optical speckle field is focused onto the atomic cloud using a converging lens.\\
A numerical algorithm to generate the intensity profile of a speckle field is based on the following expression~\cite{huntley1989speckle}:
\begin{equation}
V_{\di}(x)=V_0 \biggl| F^{-1}\left[ W(\xi) F\left[\varphi\right](\xi)\right](x)\biggr|^2.
\label{speckle1}
\end{equation}
Here, the constant $V_0$ corresponds to the average intensity of the speckle field, while
we denote with
\begin{equation}
F\left[{\varphi}\right](\xi)=\int \di{x} \varphi({x}) e^{-i2\pi{\xi}{x}}
\label{speckle2}
\end{equation}
the Fourier transform of the complex field $\varphi({ x})$, whose real and imaginary part are independent random variables sampled from a gaussian distribution with zero mean and unit variance. $F^{-1}$ indicates the inverse Fourier transform. The function $W({\xi})$ is a filter defined as
\begin{equation}
W({ \xi})=\left\{  \begin{array}{cc} 1  & \;\; \mathrm{if}\;  |\xi| \leqslant w/2  \\
                                                    0  & \;\;  \mathrm{if}\;  |\xi|>  w/2  \end{array} \right.,
\label{speckle3}
\end{equation}
where $w$ is the aperture width, which depends on the details of the optical apparatus employed to create and focus the speckle field, namely the laser wavelength, the size (illuminated area) and the focal length of the lens employed for focusing.
We consider blue detuned optical fields, for which the constant $V_0$ introduced in eq.(\ref{speckle1}) is positive. 

In the numerical implementation, the gaussian random complex field $\varphi({ x})$ is defined on a discrete grid: $x_g=g\delta x$, where $\delta x=L/N_g$, $L$ is the system size, and the integer $g=0,1,\dots, N_g-1$. The number of grid points $ N_g$  shall be large, as discussed below. The continuous Fourier transform is henceforth replaced by the discrete version. Periodic boundary conditions are adopted, and the definition (\ref{speckle1}) is consistent with this choice, i.e. $V_{\di}(L)=V_{\di}(0)$.\\
For a large enough systems size $L$, the optical speckle field is self-averaging, meaning that spatial averages coincide with the average of local values over many instances of the speckle field, indicated as $\langle V_{\di}(x)\rangle_{\di}$. These instances are realized by choosing different random numbers to define the complex field  $\varphi({ x})$. 
The probability distribution of the local speckle-field intensity $V_{\mathrm{loc}}=V_{\mathrm{d}}(x)$, for any $x$, is $P(V_{\mathrm{loc}})=\exp(-V_{\mathrm{loc}}/V_0)/V_0$. It follows that, for large enough $L$, the average speckle-field intensity $\langle V_{\di}(x) \rangle_{\di}=V_0$ is equal to the standard deviation $\sqrt{\langle V_{\di}(x)^2 \rangle_{\di}-V_0^2}=V_0$.
Therefore, $V_0$ is the unique parameter that determines the amount of disorder in the system.\\
The local speckle-field intensities at different positions have statistical correlations, characterized by the following spatial autocorrelation function:
\begin{equation}
\Gamma(x)=\langle V_{\rm{d}}({ x}^\prime)V_{\rm{d}}({x}^\prime+{x})\rangle/V_0^2-1 = \left[ \sin(\pi w x)/(\pi w x)\right]^2.
\label{speckle6}
\end{equation}
One notices that the inverse of the aperture width $w$ determines the correlation length, i.e. the typical size of the speckle grains. We will indicate this length scale as $\gamma=w^{-1}$, which corresponds to the first zero of the correlation function $\Gamma(x)$. The correlation length allows one to define an energy scale, dubbed correlation energy, defined as $E_{\ci}= \hbar^2/(2m \gamma^2)$.\\
In the following we consider the system size $L=20\gamma$, with a number of grid points $N_{\gi}=1024$. Notice that with this choice one has $\delta x \ll \gamma$, so that the discretization effect is irrelevant. Furthermore, the speckle-field intensity is fixed at the moderately large value $V_0=5E_{\ci}$.
We point out that we choose to normalize the optical speckle field so that its spatial average over the finite system size $L$ exactly corresponds to $V_0$, for each individual instance, thus eliminating small fluctuations due to finite size effects.

The local intensity profile of a typical instance of optical speckle field is displayed in the upper panel of Fig.~\ref{fig1}. The continuous horizontal line indicates the average intensity $V_0$. The lower panel displays the three eigenfunctions $\phi_i(x)$, with $i=0,1,2$, corresponding to the lowest energy levels. They solve the Schr\"odinger equation $\hat{H}\phi_i(x) = e_i \phi_i(x)$ with eigenvalues $e_i$. These energy levels are indicated by the three horizontal segments in the upper panel of Fig.~\ref{fig1}. The wave functions and the corresponding energy levels are computed via a finite difference approach, employing the grid points $x_g$ defined above, using a highly accurate $11$-point finite difference formula. This makes the discretization error negligible.
One notices that the wave functions $\phi_i(x)$ have non-negligible values only in a small region of space. This is consistent with the Anderson localization phenomenon, which in one-dimensional configurations is expected to occur for any amount of disorder, as predicted by the scaling theory of Anderson localization~\cite{PhysRevLett.42.673}.\\
Clearly, the energy levels $e_i$ randomly fluctuate for different instances of the speckle field. Their probability distribution is shown in Fig.~\ref{fig2}, where the averages over many speckle-field instances $\langle e_i \rangle_{\di}$ are also indicated with vertical segments.
One notices that the probability distribution of the ground-state energy $e_0$ is slightly asymmetric, while the distributions of the excited energy levels $e_1$ and $e_2$ appear to be essentially symmetric. Other properties of quantum particles in an optical speckle field, such as the density of states, have been investigated in Refs.~\cite{falco2010density,prat2016semiclassical}.\\
%

%

\section{Training and testing the artificial neural network}
\label{secgfmc}

The first step in a supervised machine learning study consists in choosing how to represent the system instances. 
One has to choose $N_f$ real values that describe the system, all together constituting the so-called  features vector.
One natural choice would consist in choosing the speckle field values $V_{\di}(x_g)$ on the $N_g$ points of the spatial grid defined in Sec.~\ref{secspeckle}. Indeed, if the grid is fine enough these values fully define the system Hamiltonian. However, since $N_g$ has to be large, this choice leads to a pretty large features vector,  making the training of a deep neural network with many neurons and many layers rather computational expensive. This approach was in fact adopted in a recent related article~\cite{mills2017deep}. The problem of the large feature vector was circumvented by employing so-called convolutional neural networks. In such networks the connectivity structure is limited. This reduces the number of parameters to be optimized, making the training more computationally affordable. The connectivity structure is in fact designed so that the network can recognize the spatial structures in the feature vector, somehow automatically extracting the relevant details from a large feature space.\\
In this article we adopt a different strategy. The definition of the optical speckle field in eq.~\ref{speckle1} and the structure of the spatial correlations described in Sec.~\ref{secspeckle} suggest that one can construct a more compact system representation by switching to the Fourier space. In fact, it is easy to show that the (discrete) Fourier transform of the speckle field $F\left[V_{\di}\right](\xi)$ has a finite support, limited to the interval  $\xi \in \left[-w:w\right]$. This limits the number of nonzero Fourier components. Since the Fourier grid spacing is $\delta \xi=1/L$, one expects to have $42$ nonzero (complex) Fourier components for our choice of system size $L=20\gamma$. One should also consider that the Fourier transform of a real signal has the symmetry $F\left[V_{\di}\right](-\xi)={F\left[V_{\di}\right](\xi)}^{\ast}$. This further limits the number of nonzero independent variables, leaving us with a feature vector with only $N_f=42$ (real) components.
In Fig.~\ref{fig3} we plot the average over many speckle-field instances of the absolute value of the real and imaginary parts of the Fourier components $F\left[V_{\di}\right](\xi)$. Only the positive semiaxis $\xi\geqslant 0$ is considered, due to the symmetry mentioned above. It should also be pointed out that due to the choice of normalization discussed in Sec.~\ref{secspeckle}, the real part of the Fourier transform at $\xi=0$ is fixed at $\mathrm{Re}\left\{F\left[V_{\di}\right](0)\right\}=5E_{\ci}$, for each individual speckle-field instance; also, the imaginary part is fixed at $\mathrm{Im}\left\{F\left[V_{\di}\right](0)\right\}=0$. This reduces the number of active features to $40$. Still, we include all $N_f=42$ components in the feature vector, in view of future studies extended to speckle fields with varying intensities. In fact, the inactive features do not play any role in the training of the neural network.\\
In supervised machine learning studies it is sometimes convenient  to normalize the components of the feature vector so that they have the same minimum and maximum values, or the same mean and standard deviation. This improves the efficiency in those cases in which the bare (non-normalized) feature values vary over scales that differ by several orders of magnitude. However, as can be evinced by the plot of their standard deviations (denoted $\sigma [\mathrm{Re}\left\{F[V_d](\xi)\right\}]$ and $\sigma [\mathrm{Im}\left\{F[V_d](\xi)\right\}]$) in Fig.~\ref{fig3}, the Fourier components of the speckle field differ at most by a factor of $\sim 4$. Therefore a normalization procedure is not required here.

Our plan is to train the neural network to predict the three lowest energy levels of a quantum particle in a speckle field. 
We generate a large number of speckle-field instances (we indicate this number with $N_t$) using different random numbers, as discussed in Sec.~\ref{secspeckle}.  Their energy levels are computed via the finite difference approach (see Sec.~\ref{secspeckle}). The target value is either the ground-state energy, or the first excited energy level, or the second energy level. Actually, for mere convenience, we consider the shifted energy levels $y=\epsilon_i=e_i - \langle e_i \rangle_{\di}$ (with $i=0,1,2$), so that the target values have zero mean when averaged over many speckle-field instances. Each instance is represented by the feature vector $\bold{f}=(f_1,f_2,\dots,f_{N_f})$, namely the $N_f=42$ values taken from the nonzero Fourier components described above, and the target value $y$ (ground state, first excited state, or second excited state) to be learned.
%

The statistical model we consider is a feed-forward artificial neural network, as implemented in the multi-layer perceptron regressor of the python scikit-learn library~\cite{scikit-learn}. 
This neural network includes various layers with a specified number of neurons. The leftmost layer is the input layer. It includes $N_f$ neurons, each representing one of the features values. Next, there is a tunable number of hidden layers; we indicate this number as $N_l$. This is one of the details of the statistical model that will be analyzed. The number of neurons in the hidden layers, indicated as $N_n$, can also be tuned~\cite{NoteA}. This is the second relevant detail of the statistical model to be analyzed. The input layer and the hidden layers also include a bias term. The rightmost layer is the output layer, and includes one neuron only. Each neuron $h=1,\dots,N_n$ in the hidden layer $l=1,\dots,N_l$ takes a value $a_h^l$ obtained by evaluating the so-called activation function, denoted by $g(\cdot)$, on the weighted sum $\sum_j w_{h,j}^l a_{j}^{l-1}$ of the values of the neurons in the previous layer $l-1$, adding also the bias term $b_l$, leading to: $a_h^l=g(\sum_j w_{h,j}^l a_{j}^{l-1} + b_l)$. The index $j$ labels neurons in the previous layer, so that $j=1,\dots,N_f$ when $l=1$ and $j=1,\dots,N_n$ when $l>1$. The coefficients $w_{h,j}^l$ are the weights between layer $l$ and layer $l-1$, with $l=1,\dots,N_l+1$. They represent the model parameters that have to be optimized during the learning process, together with the bias terms $b_l$. 
The neuron of the output layer (corresponding to the index $l=N_l+1$) also performs the weighted sum with bias, but the activation function is here just the identity function. 
Taking, as an illustrative example, a neural network with one hidden layer and one neuron in the hidden layer, the learning function would be $F(\bold{f}) = w_{1,1}^2 g(\sum_j w_{1,j}^{1} f_j +b_1)+b_2$.
Different choices for the activation function $g(x)$ of the hidden neurons are possible, including, e.g., the identity, the hyperbolic tangent, and the rectified linear unit function, defined as $g(x) = \mathrm{max}(0,x)$. In this article, we adopt the latter function. A preliminary analysis has shown that other suitable  choices perform quite poorly.

The learning process consists in optimizing the model parameters $w_{h,j}^l$ and $b_l$ so that the function values $F(\bold{f}_t)$ closely approximate the target values $y_t$. Here, the index $t=1,\dots,N_t$ labels the instances in the training set.
The optimization algorithm is designed to minimize the loss function $L(\bold{W})=\frac{1}{2}\sum_t \left(F(\bold{f}_t)-y_t)\right)^2 + \frac{1}{2} \alpha \| \bold{W}\|_2$, where the second term is the regularization and is introduced to penalize complex models with large coefficients. It is computed with the L2-norm, indicated as $\|\cdot\|_2$, of the vector $\bold{W}$, which includes all weight coefficients. The regularization is useful to avoid overfitting, the situation in which the target values of the training instances are accurately reproduced, but the neural network fails to correctly predict the target values of previously unseen instances.
The magnitude of the regularization term can be tuned by varying the (positive) regularization parameter $\alpha$. Typically, large values of  $\alpha$ are required when the training set is small (if the neural network has many layers and many hidden neurons), while small (or even vanishing) values of $\alpha$ can be used if the training set is sufficiently large. The role of this parameter is another aspect that will be analyzed below.
The optimization is performed using the Adam algorithm~\cite{kingma2014adam}, an improved variant of the stochastic gradient descent method, which is readily implemented in the scikit-learn library, and proves  to perform better than the other available options for our problem.
The tolerance parameter in the multi-layer perceptron regressor is set to $10^{-10}$, providing a large parameter for the maximum number of iteration so that convergence is always reached. All other parameters of the multi-layer perceptron regressor are left at their default values.\\

\begin{figure}[t]
\begin{center}
\includegraphics[width=1.0\columnwidth]{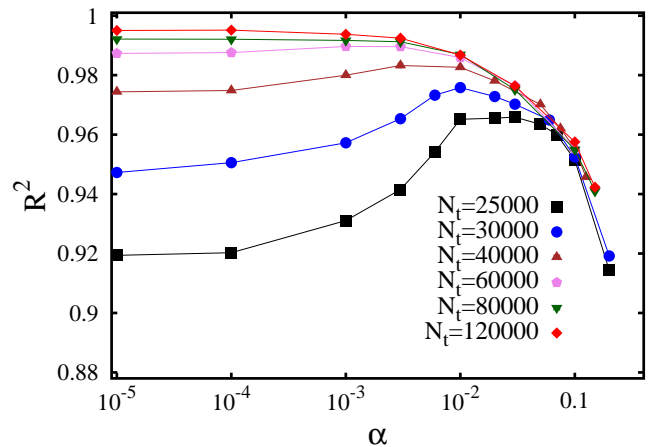}
\caption{(color online). 
Coefficient of determination $R^2$ as a function of the regularization parameter $\alpha$. The different datasets correspond to different training set sizes $N_t$. The neural network has N$_l=3$ layers and $N_n=150$ neurons per layer.
}
\label{fig4}
\end{center}
\end{figure}
%

%
In the following, we evaluate the performance of the trained neural network in predicting the energy levels of a set of $N_p=40000$ speckle-field instances not included in the training set. As a figure of merit, we consider the coefficient of determination, typically denoted with $R^2$, defined in the general case as:
\beq
 R^2=1- \frac{ \sum_{p=1}^{N_p} \left(F(\bold{f})-y_p\right)^2} {  \sum_{p=1}^{N_p} \left(y_p-\bar{y}\right)^2},
\eeq
where $\bar{y}=\frac{1}{N_p}\sum_{p=1}^{N_p} y_p$ is the average of the target values in the test set, which is essentially zero here due to the use of shifted energy levels.
A perfectly accurate statistical model which exactly predicts the target values of all the instances in the test set would yield a coefficient of determination equal to $R^2=1$. For example, a constant function which produces (only) the correct average of the test set target values, but (clearly) completely fails to reproduce their fluctuations, would instead correspond to the score $R^2=0$. Notice that the coefficient of determination could in principle be negative in the case of an extremely inaccurate statistical model (in fact, $R^2$  is not the square of a real number).
All $R^2$ scores reported in the following have been obtained as the average over $5$ to $15$ repetitions of the training of the neural network, initializing the random number generator used by the multi-layer perceptron regressor of the scikit-learn library with different seed numbers. The estimated standard deviation of the average is used to define the error bar displayed in the plots. This error bar accounts for the fluctuations due the (possibly) different local minima identified by the optimization algorithm.
%
%
\begin{figure}
\begin{center}
\includegraphics[width=1.0\columnwidth]{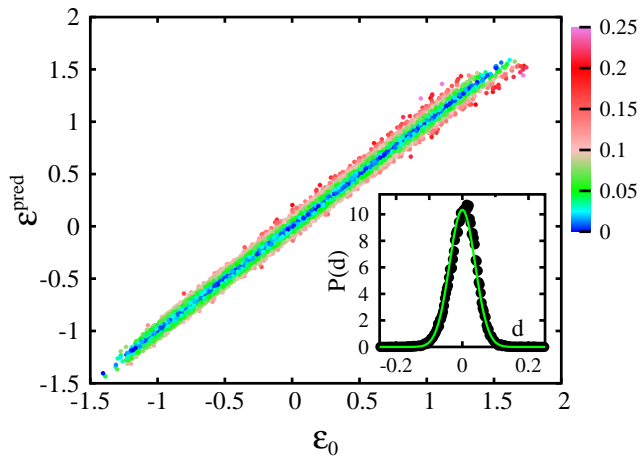}
\caption{(color online).  
Main panel: shifted energy levels predicted by the neural network $\epsilon_{\mathrm{pred}}$ as a function of the (shifted) exact values $\epsilon_0$ determined via the finite difference technique. The color scale indicates the absolute value of the discrepancy $d=\epsilon_{\mathrm{pred}}-\epsilon_0$.
Inset: probability distribution of the discrepancy $d$. The (green) continuous curve is a Gaussian fitting function 
$P(d)=  \frac{1}{\sqrt{2\pi}\sigma}  \exp\left[-d^2/(2\sigma^2)\right]$, with the fitting parameter $\sigma \cong 0.039E_{\ci}$.
}
\label{fig5}
\end{center}
\end{figure}

The first aspect of the machine learning process we analyze is the role of the regularization parameter $\alpha$. A neural network with $N_l=3$ hidden layers and $N_h=150$ neurons per hidden layer is considered for this analysis, testing how accurately it predicts the (shifted) ground state energies of the $N_p=40000$ instances of the test set. Fig.~\ref{fig4} shows the $R^2$ scores as a function of the regularization parameter, for different sizes of the training set $N_t$. One notices that for the smallest training set with $N_t=25000$ instances the optimal result is obtained with a significantly large regularization parameter, namely $\alpha \approx 0.03$. This indicates that without regularization this training set would be too small to avoid overfitting. Instead, the largest training sets provide the highest $R^2$ scores with vanishingly small $\alpha$ values, meaning that here regularization can be avoided. In fact, this neural network proves able to accurately predict the ground-state energies of the speckle-field instances, with the highest values of the coefficient of determination  $R^2$ close to 1.\\

This high accuracy can be appreciated also in the scatter plot of Fig.~\ref{fig5}, where the shifted ground-state energy $\epsilon^{\mathrm{pred}}=F(\bold{f})$ predicted by the neural network (with $N_l=3$ and $N_n=150$, as in Fig.~\ref{fig4})  is plotted versus the exact value $\epsilon_0$. Here, the training set size is $N_t=80000$, and the regularization parameter is fixed at the optimal value. The color scale indicates the absolute value of the discrepancy $d=\epsilon^{\mathrm{pred}}-e_0$. One notices that somewhat larger discrepancies occur for those speckle-field instances whose ground state energy is higher than the average. The inset of Fig.~\ref{fig5} displays its probability distribution $P(d)$. 
This distribution turns out to be well described by a gaussian fitting function with a standard deviation as small as $\sigma\cong 0.039E_c$.\\

\begin{figure}
\begin{center}
\includegraphics[width=1.0\columnwidth]{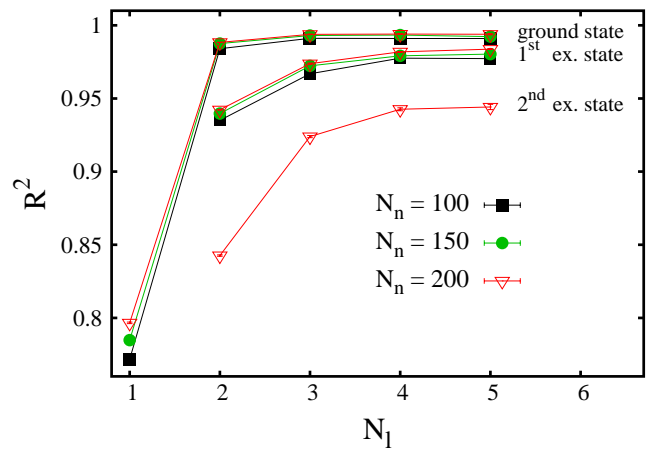}
\caption{(color online). 
Coefficient of determination $R^2$ as a function of the number of layers $N_l$ of the neural network. The upper three datasets correspond to the ground-state energy level, for three numbers of neuron per layer $N_n$. The central three datasets correspond to the first excited state. The lowest dataset corresponds to the second excited state.
}
\label{fig6}
\end{center}
\end{figure}

\begin{figure}
\begin{center}
\includegraphics[width=1.0\columnwidth]{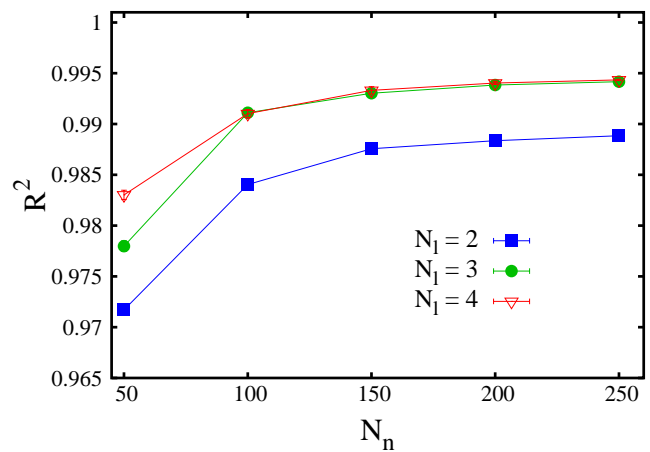}
\caption{(color online). 
Coefficient of determination $R^2$ for the prediction of ground state energy levels, as a function of the number of neurons per layer $N_n$. The three datasets correspond to different numbers of layers $N_l$.
}
\label{fig7}
\end{center}
\end{figure}

\begin{figure}
\begin{center}
\includegraphics[width=1.0\columnwidth]{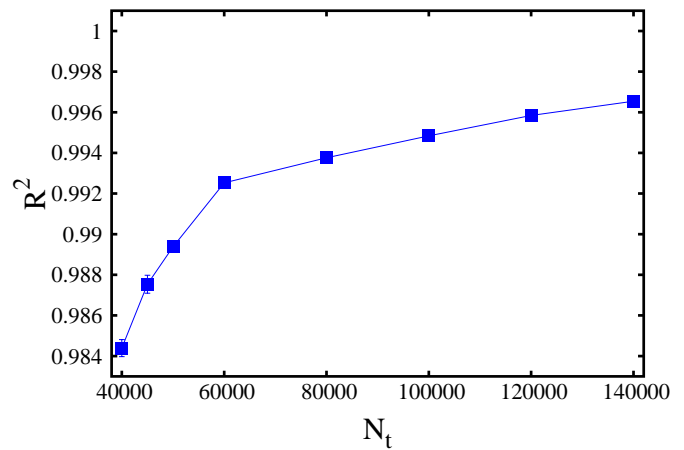}
\caption{(color online). 
Coefficient of determination $R^2$ for the prediction of ground-state energy levels as a function of the training set size $N_t$. The number of layers is $N_l=3$ and the number of neurons per layer is $N_n=200$.
}
\label{fig8}
\end{center}
\end{figure}

It is interesting to analyze how the accuracy of the neural network varies with the number of hidden layers $N_l$ and of the neuron per hidden layer $N_n$.
In Fig.~\ref{fig6} the $R^2$ scores are plotted as a function of $N_l$. The three upper datasets correspond to (shifted) ground-state energy predictions with three values of $N_h$. In Fig.~\ref{fig7} the $R^2$ scores are plotted as a function of $N_n$, for three numbers of layers $N_l$. The size of the training set is $N_t=80000$.
One notices that a neural network with only one hidden layer is not particularly accurate, with the $R^2$ score being close to $R^2\approx 0.8$. Instead, two hidden layers appear to be already sufficient to provide accurate predictions. Increasing the hidden layer number beyond $N_l=3$ does not provide a sizable accuracy improvement. The number of neurons $N_n$ plays a relevant role, too. A significant accuracy improvement occurs when the number of hidden neurons increases from $N_n=50$ to $N_n=100$. This improvement becomes less pronounced when $N_n$ is increased beyond $N_n=150$.\\
It is evident  that neural networks with $N_l> 2$ and $N_h>150$ are quite accurate statistical models to predict ground state energies; however,  their $R^2$ scores still remain close but systematically below the ideal result $R^2=1$. It is possible that a larger training set would allow one to remove even this small residual error. To address this point, we plot in Fig.~\ref{fig8} the $R^2$ score as a function of the training set size $N_t$, reaching considerably large training set sizes $N_t=140000$. The considered neural network is quite deep and wide, having $N_l=3$ hidden layers and $N_n=200$ neurons per hidden layer. 
One observes that the prediction accuracy systematically improves with $N_t$, suggesting that, with a sufficiently large training set, a deep neural network with many neurons per hidden layer can provide essentially arbitrarily accurate predictions of ground-state energies. However, it is important to point out that the training process of a deep neural network on training sets of this size becomes a computationally expensive task, at the limit of the computational resources available to us. It is worth reminding that, in general, the computational cost of training a neural network scales with the training set size, the number of features, and with the $N_l$-th power of the number of neurons per hidden layer.

We analyze also how accurately a neural network can predict the (shifted) excited state energies. In fact, in quantum many-body theory, predicting excited state energies is a more challenging 
computational problem compared to ground-state energy computations, since efficient numerical algorithms such as, e.g., quantum Monte Carlo simulations, cannot be used in general, thus demanding the use of computationally expensive techniques like exact diagonalization algorithms. It is interesting
to inspect if this greater difficulty is reflected in the process of learning to predict excited state energies from previous observations. 
$R^2$ scores corresponding to predictions of first excited state energies are displayed in Fig.~\ref{fig6}, as a function of $N_l$, for three numbers of neurons per hidden layer $N_h$. $R^2$ scores corresponding to the second excited state are displayed too, but for one $N_h$ value only.
One notices that the $R^2$ scores are lower than in the case of ground state energy predictions, in particular the results corresponding to the second excited state. Furthermore, the number of layers appears to play a more relevant role. Deeper neural networks are necessary to get close to the ideal score $R^2=1$. However, adding even more layers becomes computationally prohibitive, and is beyond the scope of this article.


\section{Conclusions}
\label{secconc}

We performed a supervised machine learning study, training a deep neural network to predict the lowest three energy levels of a quantum 
particle in a disordered external field. The quantum model we focused on is designed to describe a one-dimensional noninteracting atomic gas exposed to an optical speckle field, 
taking into account the structure of the spatial correlations of the local intensities of the random field. 
The speckle-field instances were represented using the Fourier components of the speckle field. The most relevant aspects of a supervised machine learning task have been analyzed, including the number of hidden layers in the neural network, the number of neurons in each hidden layer, the size of the training set, and the magnitude of the regularization parameter. 
Interestingly, it is found that a deep neural network with many neurons per hidden layer can provide extremely accurate prediction of ground-state energies using for training a feasible number of speckle-field instances. The predictions of excited state energies turn out to be slightly less accurate, requiring deeper neural networks to approach the optimal result.

In recent years, experiments performed with ultracold atoms have emerged as an ideal platform to perform quantum simulations of complex quantum phenomena observed also in other, less accessible and less tunable, condensed matter systems. In the long term, one can envision the use of cold-atom setups to train artificial neural network to solve problems that challenge many-body theorists, like the many-fermion problem.
In the medium term, these experiments can be employed as a testbed to develop efficient representations of instances of quantum systems for supervised machine learning tasks, as well as for testing the accuracy of different statistical models, including, e.g, the artificial neural networks, convolutional neural networks, gaussian approximation potentials, or support vector machines~\cite{mills2017deep,bartok2010gaussian}. These machine learning techniques could find use in particular in the determination of potential energy surfaces for electronic structure simulations, or even in ligand-protein affinity calculations for drug-design research. For this purpose, it is of outmost importance to understand how accurate the above mentioned statistical models can be in predicting energy levels of complex quantum many body systems. This is one of the reasons that motivated our study.\\
The Fourier representation we considered constitutes a complementary approach to other representations investigated in previous works, like the real-space grid previously employed to represent models of confined quantum particles~\cite{mills2017deep}, or the various approaches considered in the field of atomistic simulations, including, e.g, the atom-centered symmetry functions~\cite{behler2011atom}, the neighbor density, the smooth overlap of atomic positions, the Coulomb matrices (see, e.g, Ref.~\cite{behler2016perspective}), and the bag of bonds model~\cite{hansen2015machine}.
Our Fourier space representation appears to be a promising approach to describe systems characterized by external fields with spatially correlated intensities.

\acknowledgements
We acknowledge insightful discussions with Andrea De Simone. S.P. acknowledges the CINECA award under the ISCRA
initiative, for the availability of high performance computing resources and support. Partial support by the Italian
MIUR under PRIN-2015 Contract No. 2015C5SEJJ001 is also acknowledged.



\begin{thebibliography}{54}%
\makeatletter
\providecommand \@ifxundefined [1]{%
 \@ifx{#1\undefined}
}%
\providecommand \@ifnum [1]{%
 \ifnum #1\expandafter \@firstoftwo
 \else \expandafter \@secondoftwo
 \fi
}%
\providecommand \@ifx [1]{%
 \ifx #1\expandafter \@firstoftwo
 \else \expandafter \@secondoftwo
 \fi
}%
\providecommand \natexlab [1]{#1}%
\providecommand \enquote  [1]{``#1''}%
\providecommand \bibnamefont  [1]{#1}%
\providecommand \bibfnamefont [1]{#1}%
\providecommand \citenamefont [1]{#1}%
\providecommand \href@noop [0]{\@secondoftwo}%
\providecommand \href [0]{\begingroup \@sanitize@url \@href}%
\providecommand \@href[1]{\@@startlink{#1}\@@href}%
\providecommand \@@href[1]{\endgroup#1\@@endlink}%
\providecommand \@sanitize@url [0]{\catcode `\\12\catcode `\$12\catcode
  `\&12\catcode `\#12\catcode `\^12\catcode `\_12\catcode `\%12\relax}%
\providecommand \@@startlink[1]{}%
\providecommand \@@endlink[0]{}%
\providecommand \url  [0]{\begingroup\@sanitize@url \@url }%
\providecommand \@url [1]{\endgroup\@href {#1}{\urlprefix }}%
\providecommand \urlprefix  [0]{URL }%
\providecommand \Eprint [0]{\href }%
\providecommand \doibase [0]{http://dx.doi.org/}%
\providecommand \selectlanguage [0]{\@gobble}%
\providecommand \bibinfo  [0]{\@secondoftwo}%
\providecommand \bibfield  [0]{\@secondoftwo}%
\providecommand \translation [1]{[#1]}%
\providecommand \BibitemOpen [0]{}%
\providecommand \bibitemStop [0]{}%
\providecommand \bibitemNoStop [0]{.\EOS\space}%
\providecommand \EOS [0]{\spacefactor3000\relax}%
\providecommand \BibitemShut  [1]{\csname bibitem#1\endcsname}%
\let\auto@bib@innerbib\@empty
\bibitem [{\citenamefont {Snyder}\ \emph {et~al.}(2012)\citenamefont {Snyder},
  \citenamefont {Rupp}, \citenamefont {Hansen}, \citenamefont {M{\"u}ller},\
  and\ \citenamefont {Burke}}]{snyder2012finding}%
  \BibitemOpen
  \bibfield  {author} {\bibinfo {author} {\bibfnamefont {J.~C.}\ \bibnamefont
  {Snyder}}, \bibinfo {author} {\bibfnamefont {M.}~\bibnamefont {Rupp}},
  \bibinfo {author} {\bibfnamefont {K.}~\bibnamefont {Hansen}}, \bibinfo
  {author} {\bibfnamefont {K.-R.}\ \bibnamefont {M{\"u}ller}}, \ and\ \bibinfo
  {author} {\bibfnamefont {K.}~\bibnamefont {Burke}},\ }\bibfield  {title}
  {\enquote {\bibinfo {title} {Finding density functionals with machine
  learning},}\ }\href@noop {} {\bibfield  {journal} {\bibinfo  {journal} {Phys.
  Rev. Lett.}\ }\textbf {\bibinfo {volume} {108}},\ \bibinfo {pages} {253002}
  (\bibinfo {year} {2012})}\BibitemShut {NoStop}%
\bibitem [{\citenamefont {Li}\ \emph {et~al.}(2016)\citenamefont {Li},
  \citenamefont {Snyder}, \citenamefont {Pelaschier}, \citenamefont {Huang},
  \citenamefont {Niranjan}, \citenamefont {Duncan}, \citenamefont {Rupp},
  \citenamefont {M{\"u}ller},\ and\ \citenamefont
  {Burke}}]{li2016understanding}%
  \BibitemOpen
  \bibfield  {author} {\bibinfo {author} {\bibfnamefont {L.}~\bibnamefont
  {Li}}, \bibinfo {author} {\bibfnamefont {J.~C.}\ \bibnamefont {Snyder}},
  \bibinfo {author} {\bibfnamefont {I.~M.}\ \bibnamefont {Pelaschier}},
  \bibinfo {author} {\bibfnamefont {J.}~\bibnamefont {Huang}}, \bibinfo
  {author} {\bibfnamefont {U.-N.}\ \bibnamefont {Niranjan}}, \bibinfo {author}
  {\bibfnamefont {P.}~\bibnamefont {Duncan}}, \bibinfo {author} {\bibfnamefont
  {M.}~\bibnamefont {Rupp}}, \bibinfo {author} {\bibfnamefont {K.-R.}\
  \bibnamefont {M{\"u}ller}}, \ and\ \bibinfo {author} {\bibfnamefont
  {K.}~\bibnamefont {Burke}},\ }\bibfield  {title} {\enquote {\bibinfo {title}
  {Understanding machine-learned density functionals},}\ }\href@noop {}
  {\bibfield  {journal} {\bibinfo  {journal} {Int. J. Quantum Chem.}\ }\textbf
  {\bibinfo {volume} {116}},\ \bibinfo {pages} {819--833} (\bibinfo {year}
  {2016})}\BibitemShut {NoStop}%
\bibitem [{\citenamefont {Brockherde}\ \emph {et~al.}(2017)\citenamefont
  {Brockherde}, \citenamefont {Vogt}, \citenamefont {Li}, \citenamefont
  {Tuckerman}, \citenamefont {Burke},\ and\ \citenamefont
  {M{\"u}ller}}]{brockherde2017bypassing}%
  \BibitemOpen
  \bibfield  {author} {\bibinfo {author} {\bibfnamefont {F.}~\bibnamefont
  {Brockherde}}, \bibinfo {author} {\bibfnamefont {L.}~\bibnamefont {Vogt}},
  \bibinfo {author} {\bibfnamefont {L.}~\bibnamefont {Li}}, \bibinfo {author}
  {\bibfnamefont {M.~E.}\ \bibnamefont {Tuckerman}}, \bibinfo {author}
  {\bibfnamefont {K.}~\bibnamefont {Burke}}, \ and\ \bibinfo {author}
  {\bibfnamefont {K.-R.}\ \bibnamefont {M{\"u}ller}},\ }\bibfield  {title}
  {\enquote {\bibinfo {title} {{Bypassing the Kohn-Sham equations with machine
  learning}},}\ }\href@noop {} {\bibfield  {journal} {\bibinfo  {journal} {Nat.
  Commun.}\ }\textbf {\bibinfo {volume} {8}},\ \bibinfo {pages} {872} (\bibinfo
  {year} {2017})}\BibitemShut {NoStop}%
\bibitem [{\citenamefont {Snyder}\ \emph {et~al.}(2013)\citenamefont {Snyder},
  \citenamefont {Rupp}, \citenamefont {Hansen}, \citenamefont {Blooston},
  \citenamefont {M{\"u}ller},\ and\ \citenamefont {Burke}}]{snyder2013orbital}%
  \BibitemOpen
  \bibfield  {author} {\bibinfo {author} {\bibfnamefont {J.~C.}\ \bibnamefont
  {Snyder}}, \bibinfo {author} {\bibfnamefont {M.}~\bibnamefont {Rupp}},
  \bibinfo {author} {\bibfnamefont {K.}~\bibnamefont {Hansen}}, \bibinfo
  {author} {\bibfnamefont {L.}~\bibnamefont {Blooston}}, \bibinfo {author}
  {\bibfnamefont {K.-R.}\ \bibnamefont {M{\"u}ller}}, \ and\ \bibinfo {author}
  {\bibfnamefont {K.}~\bibnamefont {Burke}},\ }\bibfield  {title} {\enquote
  {\bibinfo {title} {Orbital-free bond breaking via machine learning},}\
  }\href@noop {} {\bibfield  {journal} {\bibinfo  {journal} {J. Chem Phys.}\
  }\textbf {\bibinfo {volume} {139}},\ \bibinfo {pages} {224104} (\bibinfo
  {year} {2013})}\BibitemShut {NoStop}%
\bibitem [{\citenamefont {Wang}(2016)}]{wang2016discovering}%
  \BibitemOpen
  \bibfield  {author} {\bibinfo {author} {\bibfnamefont {L.}~\bibnamefont
  {Wang}},\ }\bibfield  {title} {\enquote {\bibinfo {title} {Discovering phase
  transitions with unsupervised learning},}\ }\href@noop {} {\bibfield
  {journal} {\bibinfo  {journal} {Phys. Rev. B}\ }\textbf {\bibinfo {volume}
  {94}},\ \bibinfo {pages} {195105} (\bibinfo {year} {2016})}\BibitemShut
  {NoStop}%
\bibitem [{\citenamefont {Carrasquilla}\ and\ \citenamefont
  {Melko}(2017)}]{carrasquilla2017machine}%
  \BibitemOpen
  \bibfield  {author} {\bibinfo {author} {\bibfnamefont {J.}~\bibnamefont
  {Carrasquilla}}\ and\ \bibinfo {author} {\bibfnamefont {R.~G.}\ \bibnamefont
  {Melko}},\ }\bibfield  {title} {\enquote {\bibinfo {title} {Machine learning
  phases of matter},}\ }\href@noop {} {\bibfield  {journal} {\bibinfo
  {journal} {Nat. Phys.}\ }\textbf {\bibinfo {volume} {13}},\ \bibinfo {pages}
  {431} (\bibinfo {year} {2017})}\BibitemShut {NoStop}%
\bibitem [{\citenamefont {Van~Nieuwenburg}\ \emph {et~al.}(2017)\citenamefont
  {Van~Nieuwenburg}, \citenamefont {Liu},\ and\ \citenamefont
  {Huber}}]{van2017learning}%
  \BibitemOpen
  \bibfield  {author} {\bibinfo {author} {\bibfnamefont {E.~P.}\ \bibnamefont
  {Van~Nieuwenburg}}, \bibinfo {author} {\bibfnamefont {Y.-H.}\ \bibnamefont
  {Liu}}, \ and\ \bibinfo {author} {\bibfnamefont {S.~D.}\ \bibnamefont
  {Huber}},\ }\bibfield  {title} {\enquote {\bibinfo {title} {Learning phase
  transitions by confusion},}\ }\href@noop {} {\bibfield  {journal} {\bibinfo
  {journal} {Nat. Phys.}\ }\textbf {\bibinfo {volume} {13}},\ \bibinfo {pages}
  {435} (\bibinfo {year} {2017})}\BibitemShut {NoStop}%
\bibitem [{\citenamefont {Ch'ng}\ \emph {et~al.}(2017)\citenamefont {Ch'ng},
  \citenamefont {Carrasquilla}, \citenamefont {Melko},\ and\ \citenamefont
  {Khatami}}]{ch2017machine}%
  \BibitemOpen
  \bibfield  {author} {\bibinfo {author} {\bibfnamefont {K.}~\bibnamefont
  {Ch'ng}}, \bibinfo {author} {\bibfnamefont {J.}~\bibnamefont {Carrasquilla}},
  \bibinfo {author} {\bibfnamefont {R.~G.}\ \bibnamefont {Melko}}, \ and\
  \bibinfo {author} {\bibfnamefont {E.}~\bibnamefont {Khatami}},\ }\bibfield
  {title} {\enquote {\bibinfo {title} {Machine learning phases of strongly
  correlated fermions},}\ }\href@noop {} {\bibfield  {journal} {\bibinfo
  {journal} {Phys. Rev. X}\ }\textbf {\bibinfo {volume} {7}},\ \bibinfo {pages}
  {031038} (\bibinfo {year} {2017})}\BibitemShut {NoStop}%
\bibitem [{\citenamefont {Wetzel}(2017)}]{wetzel2017unsupervised}%
  \BibitemOpen
  \bibfield  {author} {\bibinfo {author} {\bibfnamefont {S.~J.}\ \bibnamefont
  {Wetzel}},\ }\bibfield  {title} {\enquote {\bibinfo {title} {{Unsupervised
  learning of phase transitions: From principal component analysis to
  variational autoencoders}},}\ }\href@noop {} {\bibfield  {journal} {\bibinfo
  {journal} {Phys. Rev. E}\ }\textbf {\bibinfo {volume} {96}},\ \bibinfo
  {pages} {022140} (\bibinfo {year} {2017})}\BibitemShut {NoStop}%
\bibitem [{\citenamefont {Deng}\ \emph {et~al.}(2017)\citenamefont {Deng},
  \citenamefont {Li},\ and\ \citenamefont {Sarma}}]{deng2017machine}%
  \BibitemOpen
  \bibfield  {author} {\bibinfo {author} {\bibfnamefont {D.-L.}\ \bibnamefont
  {Deng}}, \bibinfo {author} {\bibfnamefont {X.}~\bibnamefont {Li}}, \ and\
  \bibinfo {author} {\bibfnamefont {S.~D.}\ \bibnamefont {Sarma}},\ }\bibfield
  {title} {\enquote {\bibinfo {title} {Machine learning topological states},}\
  }\href@noop {} {\bibfield  {journal} {\bibinfo  {journal} {Phys. Rev. B}\
  }\textbf {\bibinfo {volume} {96}},\ \bibinfo {pages} {195145} (\bibinfo
  {year} {2017})}\BibitemShut {NoStop}%
\bibitem [{\citenamefont {Ohtsuki}\ and\ \citenamefont
  {Ohtsuki}(2017)}]{ohtsuki2017deep}%
  \BibitemOpen
  \bibfield  {author} {\bibinfo {author} {\bibfnamefont {T.}~\bibnamefont
  {Ohtsuki}}\ and\ \bibinfo {author} {\bibfnamefont {T.}~\bibnamefont
  {Ohtsuki}},\ }\bibfield  {title} {\enquote {\bibinfo {title} {Deep learning
  the quantum phase transitions in random electron systems: Applications to
  three dimensions},}\ }\href@noop {} {\bibfield  {journal} {\bibinfo
  {journal} {Jour. Phys. Soc. Jap.}\ }\textbf {\bibinfo {volume} {86}},\
  \bibinfo {pages} {044708} (\bibinfo {year} {2017})}\BibitemShut {NoStop}%
\bibitem [{\citenamefont {Hansen}\ \emph {et~al.}(2013)\citenamefont {Hansen},
  \citenamefont {Montavon}, \citenamefont {Biegler}, \citenamefont {Fazli},
  \citenamefont {Rupp}, \citenamefont {Scheffler}, \citenamefont
  {Von~Lilienfeld}, \citenamefont {Tkatchenko},\ and\ \citenamefont
  {MuÌller}}]{hansen2013assessment}%
  \BibitemOpen
  \bibfield  {author} {\bibinfo {author} {\bibfnamefont {K.}~\bibnamefont
  {Hansen}}, \bibinfo {author} {\bibfnamefont {G.}~\bibnamefont {Montavon}},
  \bibinfo {author} {\bibfnamefont {F.}~\bibnamefont {Biegler}}, \bibinfo
  {author} {\bibfnamefont {S.}~\bibnamefont {Fazli}}, \bibinfo {author}
  {\bibfnamefont {M.}~\bibnamefont {Rupp}}, \bibinfo {author} {\bibfnamefont
  {M.}~\bibnamefont {Scheffler}}, \bibinfo {author} {\bibfnamefont {O.~A.}\
  \bibnamefont {Von~Lilienfeld}}, \bibinfo {author} {\bibfnamefont
  {A.}~\bibnamefont {Tkatchenko}}, \ and\ \bibinfo {author} {\bibfnamefont
  {K.-R.}\ \bibnamefont {MuÌller}},\ }\bibfield  {title} {\enquote {\bibinfo
  {title} {Assessment and validation of machine learning methods for predicting
  molecular atomization energies},}\ }\href@noop {} {\bibfield  {journal}
  {\bibinfo  {journal} {J. Chem. Theory Comput.}\ }\textbf {\bibinfo {volume}
  {9}},\ \bibinfo {pages} {3404--3419} (\bibinfo {year} {2013})}\BibitemShut
  {NoStop}%
\bibitem [{\citenamefont {Hansen}\ \emph {et~al.}(2015)\citenamefont {Hansen},
  \citenamefont {Biegler}, \citenamefont {Ramakrishnan}, \citenamefont
  {Pronobis}, \citenamefont {Von~Lilienfeld}, \citenamefont {M\"uller},\ and\
  \citenamefont {Tkatchenko}}]{hansen2015machine}%
  \BibitemOpen
  \bibfield  {author} {\bibinfo {author} {\bibfnamefont {K.}~\bibnamefont
  {Hansen}}, \bibinfo {author} {\bibfnamefont {F.}~\bibnamefont {Biegler}},
  \bibinfo {author} {\bibfnamefont {R.}~\bibnamefont {Ramakrishnan}}, \bibinfo
  {author} {\bibfnamefont {W.}~\bibnamefont {Pronobis}}, \bibinfo {author}
  {\bibfnamefont {O.~A.}\ \bibnamefont {Von~Lilienfeld}}, \bibinfo {author}
  {\bibfnamefont {K.-R.}\ \bibnamefont {M\"uller}}, \ and\ \bibinfo {author}
  {\bibfnamefont {A.}~\bibnamefont {Tkatchenko}},\ }\bibfield  {title}
  {\enquote {\bibinfo {title} {{Machine learning predictions of molecular
  properties: Accurate many-body potentials and nonlocality in chemical
  space}},}\ }\href@noop {} {\bibfield  {journal} {\bibinfo  {journal} {J.
  Phys. Chem. Lett.}\ }\textbf {\bibinfo {volume} {6}},\ \bibinfo {pages}
  {2326--2331} (\bibinfo {year} {2015})}\BibitemShut {NoStop}%
\bibitem [{\citenamefont {Sch{\"u}tt}\ \emph {et~al.}(2014)\citenamefont
  {Sch{\"u}tt}, \citenamefont {Glawe}, \citenamefont {Brockherde},
  \citenamefont {Sanna}, \citenamefont {M{\"u}ller},\ and\ \citenamefont
  {Gross}}]{schutt2014represent}%
  \BibitemOpen
  \bibfield  {author} {\bibinfo {author} {\bibfnamefont {K.}~\bibnamefont
  {Sch{\"u}tt}}, \bibinfo {author} {\bibfnamefont {H.}~\bibnamefont {Glawe}},
  \bibinfo {author} {\bibfnamefont {F.}~\bibnamefont {Brockherde}}, \bibinfo
  {author} {\bibfnamefont {A.}~\bibnamefont {Sanna}}, \bibinfo {author}
  {\bibfnamefont {K.}~\bibnamefont {M{\"u}ller}}, \ and\ \bibinfo {author}
  {\bibfnamefont {E.}~\bibnamefont {Gross}},\ }\bibfield  {title} {\enquote
  {\bibinfo {title} {{How to represent crystal structures for machine learning:
  Towards fast prediction of electronic properties}},}\ }\href@noop {}
  {\bibfield  {journal} {\bibinfo  {journal} {Physical Review B}\ }\textbf
  {\bibinfo {volume} {89}},\ \bibinfo {pages} {205118} (\bibinfo {year}
  {2014})}\BibitemShut {NoStop}%
\bibitem [{\citenamefont {Ragoza}\ \emph {et~al.}(2017)\citenamefont {Ragoza},
  \citenamefont {Hochuli}, \citenamefont {Idrobo}, \citenamefont {Sunseri},\
  and\ \citenamefont {Koes}}]{ragoza2017protein}%
  \BibitemOpen
  \bibfield  {author} {\bibinfo {author} {\bibfnamefont {M.}~\bibnamefont
  {Ragoza}}, \bibinfo {author} {\bibfnamefont {J.}~\bibnamefont {Hochuli}},
  \bibinfo {author} {\bibfnamefont {E.}~\bibnamefont {Idrobo}}, \bibinfo
  {author} {\bibfnamefont {J.}~\bibnamefont {Sunseri}}, \ and\ \bibinfo
  {author} {\bibfnamefont {D.~R.}\ \bibnamefont {Koes}},\ }\bibfield  {title}
  {\enquote {\bibinfo {title} {Protein--ligand scoring with convolutional
  neural networks},}\ }\href@noop {} {\bibfield  {journal} {\bibinfo  {journal}
  {J. Chem. Inf. Model.}\ }\textbf {\bibinfo {volume} {57}},\ \bibinfo {pages}
  {942--957} (\bibinfo {year} {2017})}\BibitemShut {NoStop}%
\bibitem [{\citenamefont {W{\'o}jcikowski}\ \emph {et~al.}(2017)\citenamefont
  {W{\'o}jcikowski}, \citenamefont {Ballester},\ and\ \citenamefont
  {Siedlecki}}]{wojcikowski2017performance}%
  \BibitemOpen
  \bibfield  {author} {\bibinfo {author} {\bibfnamefont {M.}~\bibnamefont
  {W{\'o}jcikowski}}, \bibinfo {author} {\bibfnamefont {P.~J.}\ \bibnamefont
  {Ballester}}, \ and\ \bibinfo {author} {\bibfnamefont {P.}~\bibnamefont
  {Siedlecki}},\ }\bibfield  {title} {\enquote {\bibinfo {title} {Performance
  of machine-learning scoring functions in structure-based virtual
  screening},}\ }\href@noop {} {\bibfield  {journal} {\bibinfo  {journal} {Sci.
  Rep.}\ }\textbf {\bibinfo {volume} {7}},\ \bibinfo {pages} {46710} (\bibinfo
  {year} {2017})}\BibitemShut {NoStop}%
\bibitem [{\citenamefont {Khamis}\ \emph {et~al.}(2015)\citenamefont {Khamis},
  \citenamefont {Gomaa},\ and\ \citenamefont {Ahmed}}]{khamis2015machine}%
  \BibitemOpen
  \bibfield  {author} {\bibinfo {author} {\bibfnamefont {M.~A.}\ \bibnamefont
  {Khamis}}, \bibinfo {author} {\bibfnamefont {W.}~\bibnamefont {Gomaa}}, \
  and\ \bibinfo {author} {\bibfnamefont {W.~F.}\ \bibnamefont {Ahmed}},\
  }\bibfield  {title} {\enquote {\bibinfo {title} {Machine learning in
  computational docking},}\ }\href@noop {} {\bibfield  {journal} {\bibinfo
  {journal} {Artif. Intell. Med.}\ }\textbf {\bibinfo {volume} {63}},\ \bibinfo
  {pages} {135--152} (\bibinfo {year} {2015})}\BibitemShut {NoStop}%
\bibitem [{\citenamefont {Pereira}\ \emph {et~al.}(2016)\citenamefont
  {Pereira}, \citenamefont {Caffarena},\ and\ \citenamefont {dos
  Santos}}]{pereira2016boosting}%
  \BibitemOpen
  \bibfield  {author} {\bibinfo {author} {\bibfnamefont {J.~C.}\ \bibnamefont
  {Pereira}}, \bibinfo {author} {\bibfnamefont {E.~R.}\ \bibnamefont
  {Caffarena}}, \ and\ \bibinfo {author} {\bibfnamefont {C.~N.}\ \bibnamefont
  {dos Santos}},\ }\bibfield  {title} {\enquote {\bibinfo {title} {Boosting
  docking-based virtual screening with deep learning},}\ }\href@noop {}
  {\bibfield  {journal} {\bibinfo  {journal} {J. Chem. Inf. Model.}\ }\textbf
  {\bibinfo {volume} {56}},\ \bibinfo {pages} {2495--2506} (\bibinfo {year}
  {2016})}\BibitemShut {NoStop}%
\bibitem [{\citenamefont {Mills}\ \emph {et~al.}(2017)\citenamefont {Mills},
  \citenamefont {Spanner},\ and\ \citenamefont {Tamblyn}}]{mills2017deep}%
  \BibitemOpen
  \bibfield  {author} {\bibinfo {author} {\bibfnamefont {K.}~\bibnamefont
  {Mills}}, \bibinfo {author} {\bibfnamefont {M.}~\bibnamefont {Spanner}}, \
  and\ \bibinfo {author} {\bibfnamefont {I.}~\bibnamefont {Tamblyn}},\
  }\bibfield  {title} {\enquote {\bibinfo {title} {{Deep learning and the
  Schr{\"o}dinger equation}},}\ }\href@noop {} {\bibfield  {journal} {\bibinfo
  {journal} {Phys. Rev. A}\ }\textbf {\bibinfo {volume} {96}},\ \bibinfo
  {pages} {042113} (\bibinfo {year} {2017})}\BibitemShut {NoStop}%
\bibitem [{\citenamefont {Blank}\ \emph {et~al.}(1995)\citenamefont {Blank},
  \citenamefont {Brown}, \citenamefont {Calhoun},\ and\ \citenamefont
  {Doren}}]{blank1995neural}%
  \BibitemOpen
  \bibfield  {author} {\bibinfo {author} {\bibfnamefont {T.~B.}\ \bibnamefont
  {Blank}}, \bibinfo {author} {\bibfnamefont {S.~D.}\ \bibnamefont {Brown}},
  \bibinfo {author} {\bibfnamefont {A.~W.}\ \bibnamefont {Calhoun}}, \ and\
  \bibinfo {author} {\bibfnamefont {D.~J.}\ \bibnamefont {Doren}},\ }\bibfield
  {title} {\enquote {\bibinfo {title} {Neural network models of potential
  energy surfaces},}\ }\href@noop {} {\bibfield  {journal} {\bibinfo  {journal}
  {J. Chem. Phys.}\ }\textbf {\bibinfo {volume} {103}},\ \bibinfo {pages}
  {4129--4137} (\bibinfo {year} {1995})}\BibitemShut {NoStop}%
\bibitem [{\citenamefont {Lorenz}\ \emph {et~al.}(2004)\citenamefont {Lorenz},
  \citenamefont {Gro{\ss}},\ and\ \citenamefont
  {Scheffler}}]{lorenz2004representing}%
  \BibitemOpen
  \bibfield  {author} {\bibinfo {author} {\bibfnamefont {S.}~\bibnamefont
  {Lorenz}}, \bibinfo {author} {\bibfnamefont {A.}~\bibnamefont {Gro{\ss}}}, \
  and\ \bibinfo {author} {\bibfnamefont {M.}~\bibnamefont {Scheffler}},\
  }\bibfield  {title} {\enquote {\bibinfo {title} {Representing
  high-dimensional potential-energy surfaces for reactions at surfaces by
  neural networks},}\ }\href@noop {} {\bibfield  {journal} {\bibinfo  {journal}
  {Chem. Phys. Lett.}\ }\textbf {\bibinfo {volume} {395}},\ \bibinfo {pages}
  {210--215} (\bibinfo {year} {2004})}\BibitemShut {NoStop}%
\bibitem [{\citenamefont {Behler}\ and\ \citenamefont
  {Parrinello}(2007)}]{behler2007generalized}%
  \BibitemOpen
  \bibfield  {author} {\bibinfo {author} {\bibfnamefont {J.}~\bibnamefont
  {Behler}}\ and\ \bibinfo {author} {\bibfnamefont {M.}~\bibnamefont
  {Parrinello}},\ }\bibfield  {title} {\enquote {\bibinfo {title} {Generalized
  neural-network representation of high-dimensional potential-energy
  surfaces},}\ }\href@noop {} {\bibfield  {journal} {\bibinfo  {journal} {Phys.
  Rev. Lett.}\ }\textbf {\bibinfo {volume} {98}},\ \bibinfo {pages} {146401}
  (\bibinfo {year} {2007})}\BibitemShut {NoStop}%
\bibitem [{\citenamefont {Handley}\ and\ \citenamefont
  {Popelier}(2010)}]{handley2010potential}%
  \BibitemOpen
  \bibfield  {author} {\bibinfo {author} {\bibfnamefont {C.~M.}\ \bibnamefont
  {Handley}}\ and\ \bibinfo {author} {\bibfnamefont {P.~L.}\ \bibnamefont
  {Popelier}},\ }\bibfield  {title} {\enquote {\bibinfo {title} {Potential
  energy surfaces fitted by artificial neural networks},}\ }\href@noop {}
  {\bibfield  {journal} {\bibinfo  {journal} {J. Phys. Chem. A}\ }\textbf
  {\bibinfo {volume} {114}},\ \bibinfo {pages} {3371--3383} (\bibinfo {year}
  {2010})}\BibitemShut {NoStop}%
\bibitem [{\citenamefont {Bart{\'o}k}\ \emph {et~al.}(2010)\citenamefont
  {Bart{\'o}k}, \citenamefont {Payne}, \citenamefont {Kondor},\ and\
  \citenamefont {Cs{\'a}nyi}}]{bartok2010gaussian}%
  \BibitemOpen
  \bibfield  {author} {\bibinfo {author} {\bibfnamefont {A.~P.}\ \bibnamefont
  {Bart{\'o}k}}, \bibinfo {author} {\bibfnamefont {M.~C.}\ \bibnamefont
  {Payne}}, \bibinfo {author} {\bibfnamefont {R.}~\bibnamefont {Kondor}}, \
  and\ \bibinfo {author} {\bibfnamefont {G.}~\bibnamefont {Cs{\'a}nyi}},\
  }\bibfield  {title} {\enquote {\bibinfo {title} {{Gaussian approximation
  potentials: The accuracy of quantum mechanics, without the electrons}},}\
  }\href@noop {} {\bibfield  {journal} {\bibinfo  {journal} {Phys. Rev. Lett.}\
  }\textbf {\bibinfo {volume} {104}},\ \bibinfo {pages} {136403} (\bibinfo
  {year} {2010})}\BibitemShut {NoStop}%
\bibitem [{\citenamefont {Behler}(2011{\natexlab{a}})}]{behler2011neural}%
  \BibitemOpen
  \bibfield  {author} {\bibinfo {author} {\bibfnamefont {J.}~\bibnamefont
  {Behler}},\ }\bibfield  {title} {\enquote {\bibinfo {title} {Neural network
  potential-energy surfaces in chemistry: a tool for large-scale
  simulations},}\ }\href@noop {} {\bibfield  {journal} {\bibinfo  {journal}
  {Phys. Chem. Chem. Phys.}\ }\textbf {\bibinfo {volume} {13}},\ \bibinfo
  {pages} {17930--17955} (\bibinfo {year} {2011}{\natexlab{a}})}\BibitemShut
  {NoStop}%
\bibitem [{\citenamefont {Behler}(2011{\natexlab{b}})}]{behler2011atom}%
  \BibitemOpen
  \bibfield  {author} {\bibinfo {author} {\bibfnamefont {J.}~\bibnamefont
  {Behler}},\ }\bibfield  {title} {\enquote {\bibinfo {title} {Atom-centered
  symmetry functions for constructing high-dimensional neural network
  potentials},}\ }\href@noop {} {\bibfield  {journal} {\bibinfo  {journal} {J.
  Chem. Phys.}\ }\textbf {\bibinfo {volume} {134}},\ \bibinfo {pages} {074106}
  (\bibinfo {year} {2011}{\natexlab{b}})}\BibitemShut {NoStop}%
\bibitem [{\citenamefont {LeCun}\ \emph {et~al.}(2015)\citenamefont {LeCun},
  \citenamefont {Bengio},\ and\ \citenamefont {Hinton}}]{lecun2015deep}%
  \BibitemOpen
  \bibfield  {author} {\bibinfo {author} {\bibfnamefont {Y.}~\bibnamefont
  {LeCun}}, \bibinfo {author} {\bibfnamefont {Y.}~\bibnamefont {Bengio}}, \
  and\ \bibinfo {author} {\bibfnamefont {G.}~\bibnamefont {Hinton}},\
  }\bibfield  {title} {\enquote {\bibinfo {title} {Deep learning},}\
  }\href@noop {} {\bibfield  {journal} {\bibinfo  {journal} {Nature}\ }\textbf
  {\bibinfo {volume} {521}},\ \bibinfo {pages} {436} (\bibinfo {year}
  {2015})}\BibitemShut {NoStop}%
\bibitem [{\citenamefont {Behler}(2016)}]{behler2016perspective}%
  \BibitemOpen
  \bibfield  {author} {\bibinfo {author} {\bibfnamefont {J.}~\bibnamefont
  {Behler}},\ }\bibfield  {title} {\enquote {\bibinfo {title} {{Perspective:
  Machine learning potentials for atomistic simulations}},}\ }\href@noop {}
  {\bibfield  {journal} {\bibinfo  {journal} {J. Chem. Phys.}\ }\textbf
  {\bibinfo {volume} {145}},\ \bibinfo {pages} {170901} (\bibinfo {year}
  {2016})}\BibitemShut {NoStop}%
\bibitem [{\citenamefont {Giorgini}\ \emph {et~al.}(2008)\citenamefont
  {Giorgini}, \citenamefont {Pitaevskii},\ and\ \citenamefont
  {Stringari}}]{giorgini2008theory}%
  \BibitemOpen
  \bibfield  {author} {\bibinfo {author} {\bibfnamefont {S.}~\bibnamefont
  {Giorgini}}, \bibinfo {author} {\bibfnamefont {L.~P.}\ \bibnamefont
  {Pitaevskii}}, \ and\ \bibinfo {author} {\bibfnamefont {S.}~\bibnamefont
  {Stringari}},\ }\bibfield  {title} {\enquote {\bibinfo {title} {{Theory of
  ultracold atomic Fermi gases}},}\ }\href@noop {} {\bibfield  {journal}
  {\bibinfo  {journal} {Rev. Mod. Phys.}\ }\textbf {\bibinfo {volume} {80}},\
  \bibinfo {pages} {1215} (\bibinfo {year} {2008})}\BibitemShut {NoStop}%
\bibitem [{\citenamefont {Bloch}\ \emph {et~al.}(2008)\citenamefont {Bloch},
  \citenamefont {Dalibard},\ and\ \citenamefont {Zwerger}}]{bloch2008many}%
  \BibitemOpen
  \bibfield  {author} {\bibinfo {author} {\bibfnamefont {I.}~\bibnamefont
  {Bloch}}, \bibinfo {author} {\bibfnamefont {J.}~\bibnamefont {Dalibard}}, \
  and\ \bibinfo {author} {\bibfnamefont {W.}~\bibnamefont {Zwerger}},\
  }\bibfield  {title} {\enquote {\bibinfo {title} {Many-body physics with
  ultracold gases},}\ }\href@noop {} {\bibfield  {journal} {\bibinfo  {journal}
  {Rev. Mod. Phys.}\ }\textbf {\bibinfo {volume} {80}},\ \bibinfo {pages} {885}
  (\bibinfo {year} {2008})}\BibitemShut {NoStop}%
\bibitem [{\citenamefont {Jaksch}\ and\ \citenamefont
  {Zoller}(2005)}]{jaksch2005cold}%
  \BibitemOpen
  \bibfield  {author} {\bibinfo {author} {\bibfnamefont {D.}~\bibnamefont
  {Jaksch}}\ and\ \bibinfo {author} {\bibfnamefont {P.}~\bibnamefont
  {Zoller}},\ }\bibfield  {title} {\enquote {\bibinfo {title} {{The cold atom
  Hubbard toolbox}},}\ }\href@noop {} {\bibfield  {journal} {\bibinfo
  {journal} {Ann. Phys.}\ }\textbf {\bibinfo {volume} {315}},\ \bibinfo {pages}
  {52--79} (\bibinfo {year} {2005})}\BibitemShut {NoStop}%
\bibitem [{\citenamefont {Bernien}\ \emph {et~al.}(2017)\citenamefont
  {Bernien}, \citenamefont {Schwartz}, \citenamefont {Keesling}, \citenamefont
  {Levine}, \citenamefont {Omran}, \citenamefont {Pichler}, \citenamefont
  {Choi}, \citenamefont {Zibrov}, \citenamefont {Endres}, \citenamefont
  {Greiner}, \citenamefont {Vuleti\'c},\ and\ \citenamefont
  {Lukin}}]{bernien2017probing}%
  \BibitemOpen
  \bibfield  {author} {\bibinfo {author} {\bibfnamefont {H.}~\bibnamefont
  {Bernien}}, \bibinfo {author} {\bibfnamefont {S.}~\bibnamefont {Schwartz}},
  \bibinfo {author} {\bibfnamefont {A.}~\bibnamefont {Keesling}}, \bibinfo
  {author} {\bibfnamefont {H.}~\bibnamefont {Levine}}, \bibinfo {author}
  {\bibfnamefont {A.}~\bibnamefont {Omran}}, \bibinfo {author} {\bibfnamefont
  {H.}~\bibnamefont {Pichler}}, \bibinfo {author} {\bibfnamefont
  {S.}~\bibnamefont {Choi}}, \bibinfo {author} {\bibfnamefont {A.~S.}\
  \bibnamefont {Zibrov}}, \bibinfo {author} {\bibfnamefont {M.}~\bibnamefont
  {Endres}}, \bibinfo {author} {\bibfnamefont {M.}~\bibnamefont {Greiner}},
  \bibinfo {author} {\bibfnamefont {V.}~\bibnamefont {Vuleti\'c}}, \ and\
  \bibinfo {author} {\bibfnamefont {M.~D.}\ \bibnamefont {Lukin}},\ }\bibfield
  {title} {\enquote {\bibinfo {title} {Probing many-body dynamics on a 51-atom
  quantum simulator},}\ }\href@noop {} {\bibfield  {journal} {\bibinfo
  {journal} {Nature}\ }\textbf {\bibinfo {volume} {551}},\ \bibinfo {pages}
  {579} (\bibinfo {year} {2017})}\BibitemShut {NoStop}%
\bibitem [{\citenamefont {Roati}\ \emph {et~al.}(2008)\citenamefont {Roati},
  \citenamefont {D'Errico}, \citenamefont {Fallani}, \citenamefont {Fattori},
  \citenamefont {Fort}, \citenamefont {Zaccanti}, \citenamefont {Modugno},
  \citenamefont {Modugno},\ and\ \citenamefont {Inguscio}}]{roati2008anderson}%
  \BibitemOpen
  \bibfield  {author} {\bibinfo {author} {\bibfnamefont {G.}~\bibnamefont
  {Roati}}, \bibinfo {author} {\bibfnamefont {C.}~\bibnamefont {D'Errico}},
  \bibinfo {author} {\bibfnamefont {L.}~\bibnamefont {Fallani}}, \bibinfo
  {author} {\bibfnamefont {M.}~\bibnamefont {Fattori}}, \bibinfo {author}
  {\bibfnamefont {C.}~\bibnamefont {Fort}}, \bibinfo {author} {\bibfnamefont
  {M.}~\bibnamefont {Zaccanti}}, \bibinfo {author} {\bibfnamefont
  {G.}~\bibnamefont {Modugno}}, \bibinfo {author} {\bibfnamefont
  {M.}~\bibnamefont {Modugno}}, \ and\ \bibinfo {author} {\bibfnamefont
  {M.}~\bibnamefont {Inguscio}},\ }\bibfield  {title} {\enquote {\bibinfo
  {title} {{Anderson localization of a non-interacting Bose--Einstein
  condensate}},}\ }\href@noop {} {\bibfield  {journal} {\bibinfo  {journal}
  {Nature}\ }\textbf {\bibinfo {volume} {453}},\ \bibinfo {pages} {895--898}
  (\bibinfo {year} {2008})}\BibitemShut {NoStop}%
\bibitem [{\citenamefont {Billy}\ \emph {et~al.}(2008)\citenamefont {Billy},
  \citenamefont {Josse}, \citenamefont {Zuo}, \citenamefont {Bernard},
  \citenamefont {Hambrecht}, \citenamefont {Lugan}, \citenamefont
  {Cl{\'e}ment}, \citenamefont {Sanchez-Palencia}, \citenamefont {Bouyer},\
  and\ \citenamefont {Aspect}}]{billy2008direct}%
  \BibitemOpen
  \bibfield  {author} {\bibinfo {author} {\bibfnamefont {J.}~\bibnamefont
  {Billy}}, \bibinfo {author} {\bibfnamefont {V.}~\bibnamefont {Josse}},
  \bibinfo {author} {\bibfnamefont {Z.}~\bibnamefont {Zuo}}, \bibinfo {author}
  {\bibfnamefont {A.}~\bibnamefont {Bernard}}, \bibinfo {author} {\bibfnamefont
  {B.}~\bibnamefont {Hambrecht}}, \bibinfo {author} {\bibfnamefont
  {P.}~\bibnamefont {Lugan}}, \bibinfo {author} {\bibfnamefont
  {D.}~\bibnamefont {Cl{\'e}ment}}, \bibinfo {author} {\bibfnamefont
  {L.}~\bibnamefont {Sanchez-Palencia}}, \bibinfo {author} {\bibfnamefont
  {P.}~\bibnamefont {Bouyer}}, \ and\ \bibinfo {author} {\bibfnamefont
  {A.}~\bibnamefont {Aspect}},\ }\bibfield  {title} {\enquote {\bibinfo {title}
  {{Direct observation of Anderson localization of matter waves in a controlled
  disorder}},}\ }\href@noop {} {\bibfield  {journal} {\bibinfo  {journal}
  {Nature}\ }\textbf {\bibinfo {volume} {453}},\ \bibinfo {pages} {891--894}
  (\bibinfo {year} {2008})}\BibitemShut {NoStop}%
\bibitem [{\citenamefont {Aspect}\ and\ \citenamefont
  {Inguscio}(2009)}]{aspect2009anderson}%
  \BibitemOpen
  \bibfield  {author} {\bibinfo {author} {\bibfnamefont {A.}~\bibnamefont
  {Aspect}}\ and\ \bibinfo {author} {\bibfnamefont {M.}~\bibnamefont
  {Inguscio}},\ }\bibfield  {title} {\enquote {\bibinfo {title} {Anderson
  localization of ultracold atoms},}\ }\href@noop {} {\bibfield  {journal}
  {\bibinfo  {journal} {Phys. Today}\ }\textbf {\bibinfo {volume} {62}},\
  \bibinfo {pages} {30--35} (\bibinfo {year} {2009})}\BibitemShut {NoStop}%
\bibitem [{\citenamefont {Kondov}\ \emph {et~al.}(2011)\citenamefont {Kondov},
  \citenamefont {McGehee}, \citenamefont {Zirbel},\ and\ \citenamefont
  {DeMarco}}]{kondov2011three}%
  \BibitemOpen
  \bibfield  {author} {\bibinfo {author} {\bibfnamefont {S.}~\bibnamefont
  {Kondov}}, \bibinfo {author} {\bibfnamefont {W.}~\bibnamefont {McGehee}},
  \bibinfo {author} {\bibfnamefont {J.}~\bibnamefont {Zirbel}}, \ and\ \bibinfo
  {author} {\bibfnamefont {B.}~\bibnamefont {DeMarco}},\ }\bibfield  {title}
  {\enquote {\bibinfo {title} {{Three-dimensional Anderson localization of
  ultracold matter}},}\ }\href@noop {} {\bibfield  {journal} {\bibinfo
  {journal} {Science}\ }\textbf {\bibinfo {volume} {334}},\ \bibinfo {pages}
  {66--68} (\bibinfo {year} {2011})}\BibitemShut {NoStop}%
\bibitem [{\citenamefont {Jendrzejewski}\ \emph {et~al.}(2012)\citenamefont
  {Jendrzejewski}, \citenamefont {Bernard}, \citenamefont {Mueller},
  \citenamefont {Cheinet}, \citenamefont {Josse}, \citenamefont {Piraud},
  \citenamefont {Pezz{\'e}}, \citenamefont {Sanchez-Palencia}, \citenamefont
  {Aspect},\ and\ \citenamefont {Bouyer}}]{jendrzejewski2012three}%
  \BibitemOpen
  \bibfield  {author} {\bibinfo {author} {\bibfnamefont {F.}~\bibnamefont
  {Jendrzejewski}}, \bibinfo {author} {\bibfnamefont {A.}~\bibnamefont
  {Bernard}}, \bibinfo {author} {\bibfnamefont {K.}~\bibnamefont {Mueller}},
  \bibinfo {author} {\bibfnamefont {P.}~\bibnamefont {Cheinet}}, \bibinfo
  {author} {\bibfnamefont {V.}~\bibnamefont {Josse}}, \bibinfo {author}
  {\bibfnamefont {M.}~\bibnamefont {Piraud}}, \bibinfo {author} {\bibfnamefont
  {L.}~\bibnamefont {Pezz{\'e}}}, \bibinfo {author} {\bibfnamefont
  {L.}~\bibnamefont {Sanchez-Palencia}}, \bibinfo {author} {\bibfnamefont
  {A.}~\bibnamefont {Aspect}}, \ and\ \bibinfo {author} {\bibfnamefont
  {P.}~\bibnamefont {Bouyer}},\ }\bibfield  {title} {\enquote {\bibinfo {title}
  {Three-dimensional localization of ultracold atoms in an optical disordered
  potential},}\ }\href@noop {} {\bibfield  {journal} {\bibinfo  {journal}
  {Nature Phys.}\ }\textbf {\bibinfo {volume} {8}},\ \bibinfo {pages}
  {398--403} (\bibinfo {year} {2012})}\BibitemShut {NoStop}%
\bibitem [{\citenamefont {Anderson}(1958)}]{anderson1958absence}%
  \BibitemOpen
  \bibfield  {author} {\bibinfo {author} {\bibfnamefont {P.~W.}\ \bibnamefont
  {Anderson}},\ }\bibfield  {title} {\enquote {\bibinfo {title} {Absence of
  diffusion in certain random lattices},}\ }\href@noop {} {\bibfield  {journal}
  {\bibinfo  {journal} {Phys. Rev.}\ }\textbf {\bibinfo {volume} {109}},\
  \bibinfo {pages} {1492} (\bibinfo {year} {1958})}\BibitemShut {NoStop}%
\bibitem [{\citenamefont {Goodman}(1975)}]{goodman1975statistical}%
  \BibitemOpen
  \bibfield  {author} {\bibinfo {author} {\bibfnamefont {J.~W.}\ \bibnamefont
  {Goodman}},\ }\bibfield  {title} {\enquote {\bibinfo {title} {Statistical
  properties of laser speckle patterns},}\ }in\ \href@noop {} {\emph {\bibinfo
  {booktitle} {Laser speckle and related phenomena}}}\ (\bibinfo  {publisher}
  {Springer},\ \bibinfo {year} {1975})\ pp.\ \bibinfo {pages}
  {9--75}\BibitemShut {NoStop}%
\bibitem [{\citenamefont {Goodman}(2007)}]{goodman2007speckle}%
  \BibitemOpen
  \bibfield  {author} {\bibinfo {author} {\bibfnamefont {J.~W.}\ \bibnamefont
  {Goodman}},\ }\href@noop {} {\emph {\bibinfo {title} {Speckle phenomena in
  optics: theory and applications}}}\ (\bibinfo  {publisher} {Roberts and
  Company Publishers},\ \bibinfo {year} {2007})\BibitemShut {NoStop}%
\bibitem [{\citenamefont {Falco}\ \emph {et~al.}(2010)\citenamefont {Falco},
  \citenamefont {Fedorenko}, \citenamefont {Giacomelli},\ and\ \citenamefont
  {Modugno}}]{falco2010density}%
  \BibitemOpen
  \bibfield  {author} {\bibinfo {author} {\bibfnamefont {G.}~\bibnamefont
  {Falco}}, \bibinfo {author} {\bibfnamefont {A.~A.}\ \bibnamefont
  {Fedorenko}}, \bibinfo {author} {\bibfnamefont {J.}~\bibnamefont
  {Giacomelli}}, \ and\ \bibinfo {author} {\bibfnamefont {M.}~\bibnamefont
  {Modugno}},\ }\bibfield  {title} {\enquote {\bibinfo {title} {Density of
  states in an optical speckle potential},}\ }\href@noop {} {\bibfield
  {journal} {\bibinfo  {journal} {Phys. Rev. A}\ }\textbf {\bibinfo {volume}
  {82}},\ \bibinfo {pages} {053405} (\bibinfo {year} {2010})}\BibitemShut
  {NoStop}%
\bibitem [{\citenamefont {Modugno}(2010)}]{modugno2010anderson}%
  \BibitemOpen
  \bibfield  {author} {\bibinfo {author} {\bibfnamefont {G.}~\bibnamefont
  {Modugno}},\ }\bibfield  {title} {\enquote {\bibinfo {title} {{Anderson
  localization in Bose--Einstein condensates}},}\ }\href@noop {} {\bibfield
  {journal} {\bibinfo  {journal} {Rep. Prog. Phys.}\ }\textbf {\bibinfo
  {volume} {73}},\ \bibinfo {pages} {102401} (\bibinfo {year}
  {2010})}\BibitemShut {NoStop}%
\bibitem [{\citenamefont {Delande}\ and\ \citenamefont
  {Orso}(2014)}]{delande2014mobility}%
  \BibitemOpen
  \bibfield  {author} {\bibinfo {author} {\bibfnamefont {D.}~\bibnamefont
  {Delande}}\ and\ \bibinfo {author} {\bibfnamefont {G.}~\bibnamefont {Orso}},\
  }\bibfield  {title} {\enquote {\bibinfo {title} {Mobility edge for cold atoms
  in laser speckle potentials},}\ }\href@noop {} {\bibfield  {journal}
  {\bibinfo  {journal} {Phys. Rev. Lett.}\ }\textbf {\bibinfo {volume} {113}},\
  \bibinfo {pages} {060601} (\bibinfo {year} {2014})}\BibitemShut {NoStop}%
\bibitem [{\citenamefont {Fratini}\ and\ \citenamefont
  {Pilati}(2015{\natexlab{a}})}]{fratini}%
  \BibitemOpen
  \bibfield  {author} {\bibinfo {author} {\bibfnamefont {E.}~\bibnamefont
  {Fratini}}\ and\ \bibinfo {author} {\bibfnamefont {S.}~\bibnamefont
  {Pilati}},\ }\bibfield  {title} {\enquote {\bibinfo {title} {Anderson
  localization of matter waves in quantum-chaos theory},}\ }\href@noop {}
  {\bibfield  {journal} {\bibinfo  {journal} {Phys. Rev. A}\ }\textbf {\bibinfo
  {volume} {91}},\ \bibinfo {pages} {061601} (\bibinfo {year}
  {2015}{\natexlab{a}})}\BibitemShut {NoStop}%
\bibitem [{\citenamefont {Fratini}\ and\ \citenamefont
  {Pilati}(2015{\natexlab{b}})}]{fratini2015anderson}%
  \BibitemOpen
  \bibfield  {author} {\bibinfo {author} {\bibfnamefont {E.}~\bibnamefont
  {Fratini}}\ and\ \bibinfo {author} {\bibfnamefont {S.}~\bibnamefont
  {Pilati}},\ }\bibfield  {title} {\enquote {\bibinfo {title} {Anderson
  localization in optical lattices with correlated disorder},}\ }\href@noop {}
  {\bibfield  {journal} {\bibinfo  {journal} {Phys. Rev. A}\ }\textbf {\bibinfo
  {volume} {92}},\ \bibinfo {pages} {063621} (\bibinfo {year}
  {2015}{\natexlab{b}})}\BibitemShut {NoStop}%
\bibitem [{\citenamefont {Izrailev}\ and\ \citenamefont
  {Krokhin}(1999)}]{PhysRevLett.82.4062}%
  \BibitemOpen
  \bibfield  {author} {\bibinfo {author} {\bibfnamefont {F.~M.}\ \bibnamefont
  {Izrailev}}\ and\ \bibinfo {author} {\bibfnamefont {A.~A.}\ \bibnamefont
  {Krokhin}},\ }\bibfield  {title} {\enquote {\bibinfo {title} {Localization
  and the mobility edge in one-dimensional potentials with correlated
  disorder},}\ }\href {\doibase 10.1103/PhysRevLett.82.4062} {\bibfield
  {journal} {\bibinfo  {journal} {Phys. Rev. Lett.}\ }\textbf {\bibinfo
  {volume} {82}},\ \bibinfo {pages} {4062--4065} (\bibinfo {year}
  {1999})}\BibitemShut {NoStop}%
\bibitem [{\citenamefont {Sanchez-Palencia}\ \emph {et~al.}(2007)\citenamefont
  {Sanchez-Palencia}, \citenamefont {Cl\'ement}, \citenamefont {Lugan},
  \citenamefont {Bouyer}, \citenamefont {Shlyapnikov},\ and\ \citenamefont
  {Aspect}}]{PhysRevLett.98.210401}%
  \BibitemOpen
  \bibfield  {author} {\bibinfo {author} {\bibfnamefont {L.}~\bibnamefont
  {Sanchez-Palencia}}, \bibinfo {author} {\bibfnamefont {D.}~\bibnamefont
  {Cl\'ement}}, \bibinfo {author} {\bibfnamefont {P.}~\bibnamefont {Lugan}},
  \bibinfo {author} {\bibfnamefont {P.}~\bibnamefont {Bouyer}}, \bibinfo
  {author} {\bibfnamefont {G.~V.}\ \bibnamefont {Shlyapnikov}}, \ and\ \bibinfo
  {author} {\bibfnamefont {A.}~\bibnamefont {Aspect}},\ }\bibfield  {title}
  {\enquote {\bibinfo {title} {Anderson localization of expanding
  {Bose-Einstein} condensates in random potentials},}\ }\href {\doibase
  10.1103/PhysRevLett.98.210401} {\bibfield  {journal} {\bibinfo  {journal}
  {Phys. Rev. Lett.}\ }\textbf {\bibinfo {volume} {98}},\ \bibinfo {pages}
  {210401} (\bibinfo {year} {2007})}\BibitemShut {NoStop}%
\bibitem [{\citenamefont {Lugan}\ \emph {et~al.}(2009)\citenamefont {Lugan},
  \citenamefont {Aspect}, \citenamefont {Sanchez-Palencia}, \citenamefont
  {Delande}, \citenamefont {Gr\'emaud}, \citenamefont {M\"uller},\ and\
  \citenamefont {Miniatura}}]{PhysRevA.80.023605}%
  \BibitemOpen
  \bibfield  {author} {\bibinfo {author} {\bibfnamefont {P.}~\bibnamefont
  {Lugan}}, \bibinfo {author} {\bibfnamefont {A.}~\bibnamefont {Aspect}},
  \bibinfo {author} {\bibfnamefont {L.}~\bibnamefont {Sanchez-Palencia}},
  \bibinfo {author} {\bibfnamefont {D.}~\bibnamefont {Delande}}, \bibinfo
  {author} {\bibfnamefont {B.}~\bibnamefont {Gr\'emaud}}, \bibinfo {author}
  {\bibfnamefont {C.~A.}\ \bibnamefont {M\"uller}}, \ and\ \bibinfo {author}
  {\bibfnamefont {C.}~\bibnamefont {Miniatura}},\ }\bibfield  {title} {\enquote
  {\bibinfo {title} {One-dimensional {Anderson} localization in certain
  correlated random potentials},}\ }\href {\doibase 10.1103/PhysRevA.80.023605}
  {\bibfield  {journal} {\bibinfo  {journal} {Phys. Rev. A}\ }\textbf {\bibinfo
  {volume} {80}},\ \bibinfo {pages} {023605} (\bibinfo {year}
  {2009})}\BibitemShut {NoStop}%
\bibitem [{\citenamefont {Huntley}(1989)}]{huntley1989speckle}%
  \BibitemOpen
  \bibfield  {author} {\bibinfo {author} {\bibfnamefont {J.}~\bibnamefont
  {Huntley}},\ }\bibfield  {title} {\enquote {\bibinfo {title} {Speckle
  photography fringe analysis: assessment of current algorithms},}\ }\href@noop
  {} {\bibfield  {journal} {\bibinfo  {journal} {Appl. Opt.}\ }\textbf
  {\bibinfo {volume} {28}},\ \bibinfo {pages} {4316--4322} (\bibinfo {year}
  {1989})}\BibitemShut {NoStop}%
\bibitem [{\citenamefont {Abrahams}\ \emph {et~al.}(1979)\citenamefont
  {Abrahams}, \citenamefont {Anderson}, \citenamefont {Licciardello},\ and\
  \citenamefont {Ramakrishnan}}]{PhysRevLett.42.673}%
  \BibitemOpen
  \bibfield  {author} {\bibinfo {author} {\bibfnamefont {E.}~\bibnamefont
  {Abrahams}}, \bibinfo {author} {\bibfnamefont {P.~W.}\ \bibnamefont
  {Anderson}}, \bibinfo {author} {\bibfnamefont {D.~C.}\ \bibnamefont
  {Licciardello}}, \ and\ \bibinfo {author} {\bibfnamefont {T.~V.}\
  \bibnamefont {Ramakrishnan}},\ }\bibfield  {title} {\enquote {\bibinfo
  {title} {Scaling theory of localization: Absence of quantum diffusion in two
  dimensions},}\ }\href {\doibase 10.1103/PhysRevLett.42.673} {\bibfield
  {journal} {\bibinfo  {journal} {Phys. Rev. Lett.}\ }\textbf {\bibinfo
  {volume} {42}},\ \bibinfo {pages} {673--676} (\bibinfo {year}
  {1979})}\BibitemShut {NoStop}%
\bibitem [{\citenamefont {Prat}\ \emph {et~al.}(2016)\citenamefont {Prat},
  \citenamefont {Cherroret},\ and\ \citenamefont
  {Delande}}]{prat2016semiclassical}%
  \BibitemOpen
  \bibfield  {author} {\bibinfo {author} {\bibfnamefont {T.}~\bibnamefont
  {Prat}}, \bibinfo {author} {\bibfnamefont {N.}~\bibnamefont {Cherroret}}, \
  and\ \bibinfo {author} {\bibfnamefont {D.}~\bibnamefont {Delande}},\
  }\bibfield  {title} {\enquote {\bibinfo {title} {Semiclassical spectral
  function and density of states in speckle potentials},}\ }\href@noop {}
  {\bibfield  {journal} {\bibinfo  {journal} {Physical Review A}\ }\textbf
  {\bibinfo {volume} {94}},\ \bibinfo {pages} {022114} (\bibinfo {year}
  {2016})}\BibitemShut {NoStop}%
\bibitem [{\citenamefont {Pedregosa}\ \emph {et~al.}(2011)\citenamefont
  {Pedregosa}, \citenamefont {Varoquaux}, \citenamefont {Gramfort},
  \citenamefont {Michel}, \citenamefont {Thirion}, \citenamefont {Grisel},
  \citenamefont {Blondel}, \citenamefont {Prettenhofer}, \citenamefont {Weiss},
  \citenamefont {Dubourg}, \citenamefont {Vanderplas}, \citenamefont {Passos},
  \citenamefont {Cournapeau}, \citenamefont {Brucher}, \citenamefont {Perrot},\
  and\ \citenamefont {Duchesnay}}]{scikit-learn}%
  \BibitemOpen
  \bibfield  {author} {\bibinfo {author} {\bibfnamefont {F.}~\bibnamefont
  {Pedregosa}}, \bibinfo {author} {\bibfnamefont {G.}~\bibnamefont
  {Varoquaux}}, \bibinfo {author} {\bibfnamefont {A.}~\bibnamefont {Gramfort}},
  \bibinfo {author} {\bibfnamefont {V.}~\bibnamefont {Michel}}, \bibinfo
  {author} {\bibfnamefont {B.}~\bibnamefont {Thirion}}, \bibinfo {author}
  {\bibfnamefont {O.}~\bibnamefont {Grisel}}, \bibinfo {author} {\bibfnamefont
  {M.}~\bibnamefont {Blondel}}, \bibinfo {author} {\bibfnamefont
  {P.}~\bibnamefont {Prettenhofer}}, \bibinfo {author} {\bibfnamefont
  {R.}~\bibnamefont {Weiss}}, \bibinfo {author} {\bibfnamefont
  {V.}~\bibnamefont {Dubourg}}, \bibinfo {author} {\bibfnamefont
  {J.}~\bibnamefont {Vanderplas}}, \bibinfo {author} {\bibfnamefont
  {A.}~\bibnamefont {Passos}}, \bibinfo {author} {\bibfnamefont
  {D.}~\bibnamefont {Cournapeau}}, \bibinfo {author} {\bibfnamefont
  {M.}~\bibnamefont {Brucher}}, \bibinfo {author} {\bibfnamefont
  {M.}~\bibnamefont {Perrot}}, \ and\ \bibinfo {author} {\bibfnamefont
  {E.}~\bibnamefont {Duchesnay}},\ }\bibfield  {title} {\enquote {\bibinfo
  {title} {Scikit-learn: Machine learning in {P}ython},}\ }\href@noop {}
  {\bibfield  {journal} {\bibinfo  {journal} {J. Mach. Learn. Res.}\ }\textbf
  {\bibinfo {volume} {12}},\ \bibinfo {pages} {2825--2830} (\bibinfo {year}
  {2011})}\BibitemShut {NoStop}%
\bibitem [{Not()}]{NoteA}%
  \BibitemOpen
  \href@noop {} {}\bibinfo {note} {Different hidden layers could have different
  number of neurons. However, in this article we choose to have the same number
  of neurons $N_n$ in all hidden layers}\BibitemShut {NoStop}%
\bibitem [{\citenamefont {Kingma}\ and\ \citenamefont
  {Ba}(2014)}]{kingma2014adam}%
  \BibitemOpen
  \bibfield  {author} {\bibinfo {author} {\bibfnamefont {D.~P.}\ \bibnamefont
  {Kingma}}\ and\ \bibinfo {author} {\bibfnamefont {J.}~\bibnamefont {Ba}},\
  }\href@noop {} {\enquote {\bibinfo {title} {{Adam: A method for stochastic
  optimization}},}\ } (\bibinfo {year} {2014}),\ \Eprint
  {http://arxiv.org/abs/1412.6980} {arXiv:1412.6980} \BibitemShut {NoStop}%
\end{thebibliography}

%

\end{document}